\documentclass[a4paper,10pt]{article}

\usepackage{graphicx}
\usepackage{hyperref}

\usepackage{tabularx}

\usepackage{multirow}

\usepackage{amsmath}

\usepackage{natbib}

\usepackage{footmisc}

\usepackage{authblk}

\date{Apr. 18, 2010\footnote{This article was originally submitted to the Journal of Artificial Societies and Social Simulation (JASSS) in 2010 and was accepted for publication on Apr. 18th, 2010. Yet for personal reasons independent of the research content of this article, the publication process never reached its end. As the reviewers are kind enough to remember this article and raise questions about it many years later, I release it today in Feb. 2020 out of its original scientific context.}}

\title{Small world is not enough: Criteria for network choice and conclusiveness of simulations}
\author{Samuel Thiriot\\ \texttt{samuel.thiriot@res-ear.ch} \\ IRIT, Toulouse, France}

\newcommand{\ifthenelsehtml}[2]{#2}

\begin{document}

\bibliographystyle{jtbnew}
\setcitestyle{authoryear,round,aysep={}}

% python jasssTexPreprocessor.py different_small_world.tex
% 
%  latex -interaction=nonstopmode different_small_world_PRE.tex 
%   bibtex different_small_world_PRE
% 
% 
% 
% 

\ifthenelsehtml{
	\def\hautPrTrois{65mm}
	\def\hautPrTroisCarres{65mm}
	\def\largPrExReseauxA{60mm} 
	\def\largPrExReseauxB{60mm}
	\def\largPrExReseauxC{60mm}
	\def\largPrSchemas{200mm}
	\def\mosaicHa{70mm}
	\def\mosaicHb{60mm}
	\def\mosaicHc{60mm}
	\def\mosaicMa{65mm}
	\def\mosaicMb{60mm}
	\def\mosaicMc{60mm}
	\def\mosaicBa{75mm}
	\def\mosaicBb{60mm}
	\def\mosaicBc{60mm}
	\DeclareGraphicsExtensions{.png,.jpg,.pdf,.eps}
	% \DeclareGraphicsExtensions{.eps,.png,.jpg,.pdf}
} {
	\def\hautPrTrois{0.26\textwidth}
	\def\hautPrTroisCarres{0.32\textwidth}
	\def\largPrExReseauxA{0.29\textwidth}
	\def\largPrExReseauxB{0.35\textwidth}
	\def\largPrExReseauxC{0.31\textwidth}	
	\def\largPrSchemas{0.8\textwidth}
	\def\mosaicHa{0.31\textwidth}
	\def\mosaicHb{0.29\textwidth}
	\def\mosaicHc{0.29\textwidth}
	\def\mosaicMa{0.31\textwidth}
	\def\mosaicMb{0.29\textwidth}
	\def\mosaicMc{0.29\textwidth}
	\def\mosaicBa{0.32\textwidth}
	\def\mosaicBb{0.29\textwidth}
	\def\mosaicBc{0.29\textwidth}
	\DeclareGraphicsExtensions{.pdf,.eps,.png,.jpg}
}

\maketitle

\begin{abstract}
Most agent-based models include a social network that describes the structure of interactions within the artificial population. Because of the dramatic impact of this structure on the simulated dynamics, modellers create this network for it to match criteria of plausibility (e.g. the small-world property). Networks are actually created by one network generator compliant with these criteria, like the Watts-Strogatz algorithm in the case of small-world networks. However, this practice comes to study the model's dynamics over the specific networks generated by one algorithm instead of the dynamics over the class of networks of interest, possibly inducing a strong bias in results. We identify three problematics related to this bias: (i) representativity of a network generator to a class of networks, (ii) conclusiveness of simulations over a class of networks and (iii) the gain in conclusiveness when refining the criteria for network choice. We propose an experimental protocol and instanciate it on small-world networks for epidemics, opinion and culture dynamics. We show that (i) Watts-Strogatz networks are not representative of small-world networks (ii) simulation results over small-worlds are arguably inconclusive, and (iii) even small-world networks having the same size, density, transivity and average path length do not lead to coherent results. Beyond questionning the relevance of simulation results obtained from artificial networks, this research also constitute one more argument for the exploration of other approaches that are not solely focused on networks' statistical properties.


\end{abstract}

% 
% Choice of a network
% Dramatic influence
% Right critera
% Two mains: (1) plausibility (2) influence on collective dynamics

\section{Introduction}

\subsection{Foreword}

%\paragraph{}
Lets us take the case of a modeler willing to study the Influenza epidemics with the help of an agent-based model. Following common practices, he will use an interaction network to describe the structure of interactions within the artificial population, which is known to have a dramatic impact on simulation results. Because simulation results will be used for crucial decision-making like the definition of a vaccination strategy, the modeler will try to use a plausible network. A quick review of common practices will teach him that (i) many statistical indicators were recently proposed, but their genericity was not yet demonstrated (ii) real networks are proved to comply with the small-world phenomenon (iii) agent-based modelers classically generate small-world networks using the famous Watts-Strogatz algorithm. After a (hopefully) careful parameter setting for his epidemic model, he will thus explore its dynamics over Watts-Strogatz networks and submit his/er recommendations for fighthing Influenza. However, this practice suffer a potential flaw: our modeler argues of the plausibility of his results because simulations are based on a \textit{class of networks that is plausible}, while his simulations actually take place over \textit{specific networks generated by one algorithm}. In practice, there is not evidence that simulation results over the Watts-Strogatz networks are \textit{representative} of the dynamics of the same model across all the possible networks. There is even no evidence that simulations in the whole class of networks would be \textit{conclusive} at all; it is even possible that trying to reduce the space of possible networks by \textit{refining the criteria} for network choice (like a precise value of the average degree) would not lead to more conclusive results. 

\subsection{Problematics\label{indoc:problematics_intro}}

%\paragraph{}
This problematics extends to computational simulation in its whole and goes beyond the specific case of the Watts-Strogatz small-world networks. Whatever the criteria assumed to characterize real networks (small-world, skewed distribution of degree, assortativity, etc.), modelers eventually use one or two specific network generators (possibility an ad-hoc algorithm of their own) to actually create networks compliant with these criteria. We identify three potential flaws and the corresponding problematics related to this topic:
\begin{itemize}
\item Questioning the \textit{generator representativity to a class of networks} comes to evaluate whether the dynamics using the specific networks from a given generator are similar to the dynamics of the very same model over other networks of the same class. In the case of small-world networks, it comes to study whether the dynamics of a model over Watts-Strogatz networks are peculiar to these networks, or if they are similar to other small-worlds in general. If the difference between these dynamics is important, Watts-Strogatz networks cannot be said representative to the class of small-world networks; thereby simulations based on Watts-Strogatz networks only cannot be extrapolated to the possible dynamics occurring over the class of plausible networks. This problematic actually questions the relevance of any conclusion based on computational simulation over artificial networks.

\item Assessing the \textit{conclusiveness of simulations over a class of networks} comes to check whether the dynamics of a model over the whole class of networks that are assumed to be plausible (rather than networks from one sole generator) lead to conclusive or inconclusive results. In the case of small-worlds, simulations over the whole space of small-worlds networks may lead to so inconclusive results that the model would fail to play its filter role~\citep{samuel_thiriot:bib_sma_simulation:legay_1973_1} and thereby be useless. In case of inconclusiveness of simulations over a class of networks, the need to refine the criteria that define that class of networks would be proved. 

\item The \textit{criteria refinability potential} refers to the possibility to provide more precise (quantitative) criteria for network choice, such that simulation results become conclusive. In the case of small-worlds, when modelers use a plausible average degree for their networks (e.g.~\citep{samuel_thiriot:bib_sma_simulation:small_2005_1}), it comes to assume that refining the small-world criteria with a quantitative value will lead to more precise results. One more time, there is no evidence that other artificial networks having exactly the same characteristics would not lead to similar conclusions. In fact, we show in this paper that even networks having similar size, average degree, clustering and average path length may lead to inconclusive results.
\end{itemize}

\subsection{Related work\label{indoc:relatedwork}}

%\paragraph{}
Several studies already highlighted that \textit{different artificial networks lead to different dynamics}, as shown by the various reviews on this topic \citep{samuel_thiriot:bib_sma_simulation:albert_2002_1,samuel_thiriot:bib_sma_simulation:newman_2003_5,samuel_thiriot:bib_sma_simulation:boccaletti_2006_1}. For instance, epidemic dynamics were shown to differ across random networks, Watts-Strogatz small-world networks~\citep{samuel_thiriot:bib_sma_simulation:watts_1998_1,samuel_thiriot:bib_sma_simulation:newman_1999_2,samuel_thiriot:bib_sma_simulation:moore_2000_1} or Barab\'asi-Albert scale-free networks \citep{samuel_thiriot:bib_sma_simulation:pastorsatorras_2001_1,samuel_thiriot:bib_sma_simulation:may_2001_1,samuel_thiriot:bib_sma_simulation:eguiluz_2002_1}. In a similar way, opinion dynamics over networks were studied over small-worlds \citep{samuel_thiriot:bib_sma_simulation:suo_2008_1,samuel_thiriot:bib_sma_simulation:deffuant_2006_1,samuel_thiriot:bib_sma_simulation:weisbuch_2004_1,samuel_thiriot:bib_sma_simulation:amblard_2004_1} and scale-free networks \citep{samuel_thiriot:bib_sma_simulation:stauffer_2004_1}.  However, these experiments are driven in different experimental settings, with different implementation and parameters, thus failing to deliver a comprehensive look of the differences of dynamics of the very same model over various networks. 

%\paragraph{}
Few transverse experiments compared the dynamics over different networks. Dekker compared three models of organization over networks having different properties \citep{samuel_thiriot:bib_sma_simulation:dekker_2007_1}. Opinion dynamics over scale-free, small-world and real networks were compared by Cointet \& Roth \citep{samuel_thiriot:bib_sma_simulation:cointet_2007_1,samuel_thiriot:bib_sma_simulation:cointet_2007_2}; Ru also studied the impact of communities of the dynamics of opinions \citep{samuel_thiriot:bib_sma_simulation:ru_2008_1}. Klemm compared culture dynamics over various complex networks \citep{samuel_thiriot:bib_sma_simulation:klemm_2003_1}. Deffuant also highlighted the difference in dynamics when using random networks, regular lattices and small-world networks for opinion dynamics \citep{samuel_thiriot:bib_sma_simulation:deffuant_2006_1}. Crepey analyzed the influence of the characteristics of real networks, scale-free and random networks on epidemics \citep{samuel_thiriot:bib_sma_simulation:crepey_2006_1}. \textit{All these studies reveal important differences in the dynamics depending to the networks}; however, none tackled the three novel problematics listed before, that are focused on the conclusiveness and confidence of simulation in general rather than on the dynamics of one specific model.

%\paragraph{}
The search for novel indicators that characterize social networks is already on the center of the tremendous activity devoted to complex networks~\citep{samuel_thiriot:bib_sma_simulation:jackson_2008_1,samuel_thiriot:bib_sma_simulation:newman_2006_1}, leading to the definition of many novel indicators \citep{samuel_thiriot:bib_sma_simulation:costa_2007_1}, quickly followed by new network generators that mimic these properties. In another stream of research, many researchers work on the building of networks from local and plausible behaviors, arguing of the importance of network plausibility (e.g. \citep{samuel_thiriot:bib_sma_simulation:roth_2007_1}). However, as long as the limits of existing artificial networks are not clarified, the necessity of this difficult quest remains difficult to defend.

\subsection{Outline}

%\paragraph{}
We first (\ref{indoc:protocol}) introduce some formalism to clarify the problematic of defining criteria for network choice, and point out that this choice in not only based on the plausibility of networks, but rather on the possibility to obtain conclusive simulation results over the space of networks defined by these criteria. After setting up the general protocol to investigate the interplay between criteria for network choice and dynamics, we describe (\ref{indoc:protocol_WS}) the experimental setting used for the experiments focused on the small-world phenomenon (five network generators and nine samples of the space of small-world networks). Section~\ref{indoc:experiments} presents simulation results over the space of small-worlds and on networks having similar characteristics. As discussed in the last part of the paper (\ref{indoc:discussion}), these experiments demonstrate that the definition of what small-world networks are is too imprecise to lead to coherent simulation results. Moreover, simulations ran over Watts-Strogatz networks are shown not to be representative of the ones ran over other networks.

\section{Approach: Criteria, Networks \& Dynamics\label{indoc:protocol}}

\subsection{Formal problematics\label{indoc:formal}}

%\paragraph{}
The definition of criteria for assessing the plausibility of networks is usually an implicit and discursive process. In order to clarify this problematic, we start by formalizing the concepts of network spaces and their impact on dynamics. This formalism will facilitate the description of the common practices, the potential flaws pointed out in introduction, and justify the experimental setting used in this paper.

\subsubsection{Criteria \& network spaces}

\begin{figure}[ht]
\centering
\ifthenelsehtml{
\includegraphics{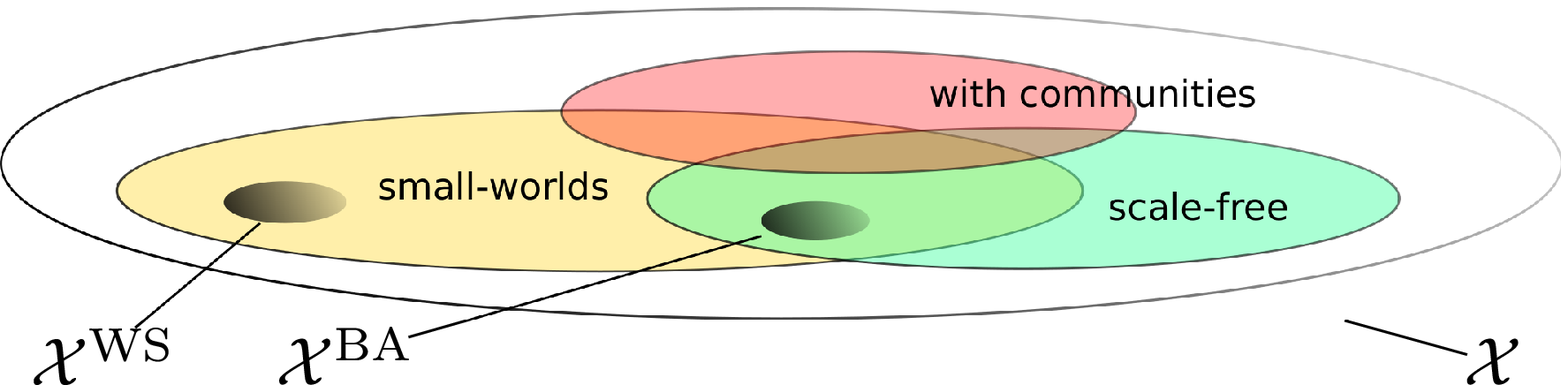}
} {
\includegraphics[width=0.6\textwidth]{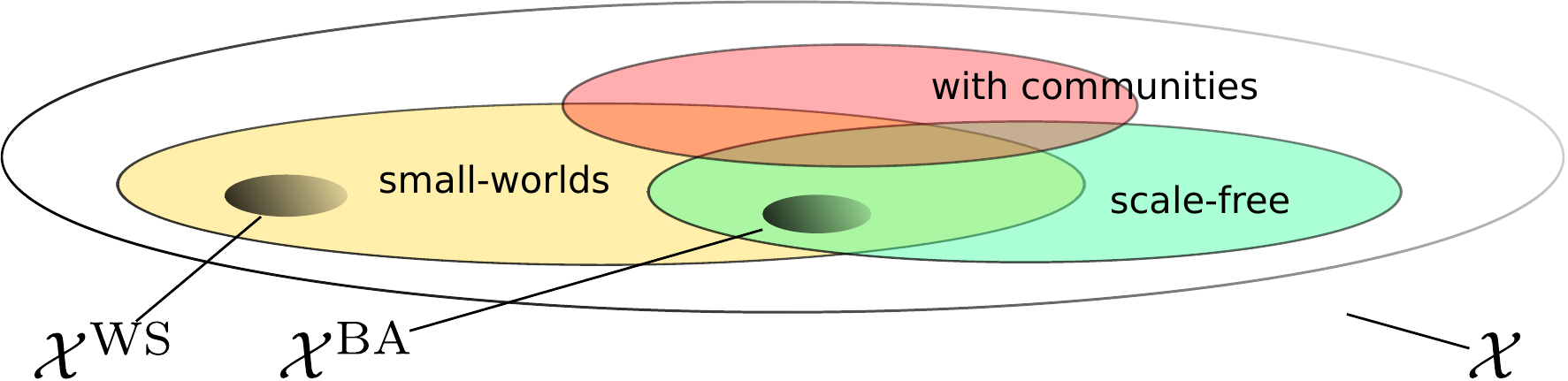}
}
\caption{Illustration of the whole space of networks $\mathcal{X}$, the subspaces of small-world, scale-free, or community-structured networks, and the specific subspaces $\mathcal{X}^{\text{WS}}$ and $\mathcal{X}^{\text{BA}}$ corresponding to the networks generated resp. by the Watts-Strogatz and the Barab\'asi-Albert algorithms}
\label{fig:conceptual_spaces}
\end{figure} 

%\paragraph{}
The problematic of defining the structure of interaction for a simulation run comes to select one network $X^{sim}$ among the space of all the possible networks $\mathcal{X}$. Following the common practices in agent-based modelling, the space $\mathcal{X}$ will here be assumed to contain only unweighted and undirected networks. The real structure of interactions $X^{real}$ that supports the real social phenomenon is unknown\footnote{Not only the real structure of interactions is unknown and unobservable, but it could also be said not to exist at all. Social networks are a metaphor \citep{samuel_thiriot:bib_psycho:breiger_2004_1} that enable us to represent and analyze the complex patterns of interactions; networks are a construction of ours rather than a real object.}: extensive data collecting is intractable because of both its prohibitive cost and the unavoidable biases in network collecting and sampling \citep{samuel_thiriot:bib_psycho:scott_2001_1,samuel_thiriot:bib_psycho:alba_1982_1,samuel_thiriot:bib_psycho:frank_1978_1}. Modelers rather use artificial networks they estimate to be plausible, because they comply with evidence on the characteristics of several real networks. Thereby modelers start, even in an unconscious and implicit way, by defining a restricted set of the properties they believe to characterize real networks. Put in a formal way, this selection comes to define \textit{criteria of plausibility} $E^{\text{plausible}}$ that restrict the space of all possible networks to the subset $\mathcal{X}^{\text{plausible}} \subset \mathcal{X}$ of networks assumed to be plausible given $E^{\text{plausible}}$. Incidentally, a subset of networks $\mathcal{X}^{\text{plausible}}$ defined by given criteria may also be name a ``class'' of networks. In practice, the \textit{criteria for network choice} $E^{\text{plausible}}$ \textit{are a set of propositions related to the statistical properties observed in real networks}; these propositions are often qualitative. For instance, when modelers argue that they ``use a Watts-Strogatz small-world network because of its compliance with both the high clustering rate and the short average path length observed in real networks'', they define criteria of network plausibility $E^{small worlds}= \{ \text{high clustering}, \text{short average path length}, \text{low density}\}$. 

% 
% 
% These observations may be thought as stylized facts available as values, or coexistence of values, of various statistical indicators. 
% 
% %\paragraph{}
% Many indicators were recently provided in the tremendous activity on complex networks: distribution of degrees, assortativity, TODO. Many of these properties were only highlighted for specific networks, while others are discussed (as the power-law distribution of degree TODO). 

\subsubsection{Random network generators}

%\paragraph{}
Once the choice of criteria to assess network plausibility is made, modelers use random network generators to create the network $X^{sim}$ that will be used for a simulation run. A random network generator $G^j$ is a generative algorithm that was built such that the networks it creates are compliant with a set of criteria of interest $E^{i}$, thus creating networks in the subspace of networks $\mathcal{X}^i$. As an example, the famous Watts-Strogatz $\beta$-model was created to reproduce the properties $E^{qWS}$ \citep{samuel_thiriot:bib_sma_simulation:watts_1998_1}. \newline
Being \textit{random} network generators, these algorithms involve a stochastic component during the generation process. A generator $G^j$ may thus be said to ``explore'' the subspace of networks $\mathcal{X}^j$. In practice, the only guarantee offered by network generators is that the network they build comply with the constraints: $\mathcal{X}^j \subset \mathcal{X}^{i}$ (e.g. the Watts-Strogatz algorithm create networks compliant with the small-world phenomenon).\newline 
It is important to note that \textit{that networks generated by $G^j$ do not cover the entire space} $\mathcal{X}^{i}$ of networks compliant with these properties. As illustrated in figure~\ref{fig:conceptual_spaces} for the space of small-world networks $\mathcal{X}^{small worlds}$, the Watts-Strogatz generator only creates networks having a Poisson-law distribution of degree, while the entire set $\mathcal{X}^{small worlds}$ also contains networks having a fat-tailed distribution of degrees or composed of many communities. The \textit{representativity problem} mentioned in introduction is precisely due to the fact that networks $\mathcal{X}^i$ generated by a given algorithm $G^i$ only constitute specific examples of the class of networks thought to be plausible $\mathcal{X}^{\text{plausible}}$ (see~\ref{indoc:protocol_WS}).  

\begin{figure}[t]
\centering
\ifthenelsehtml{
\includegraphics{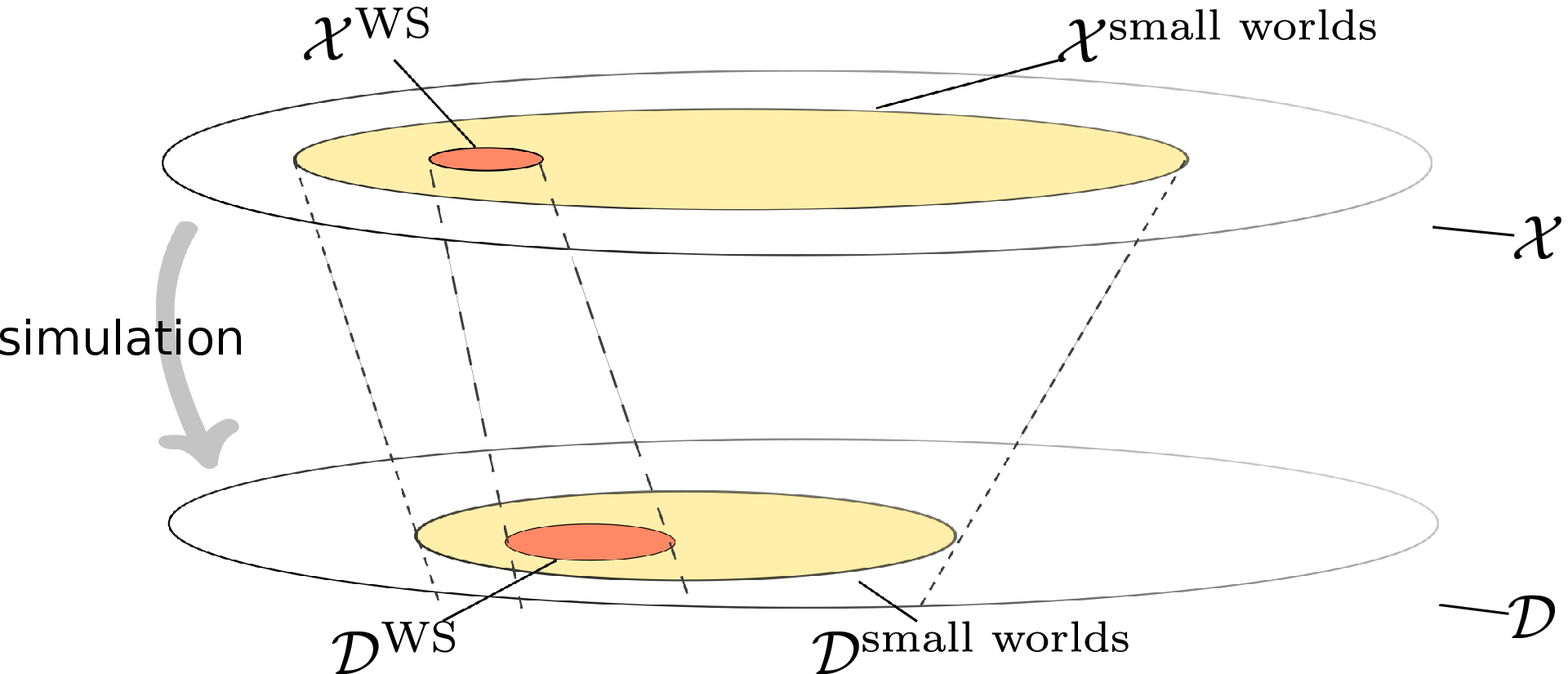}
} {
\includegraphics[width=0.6\textwidth]{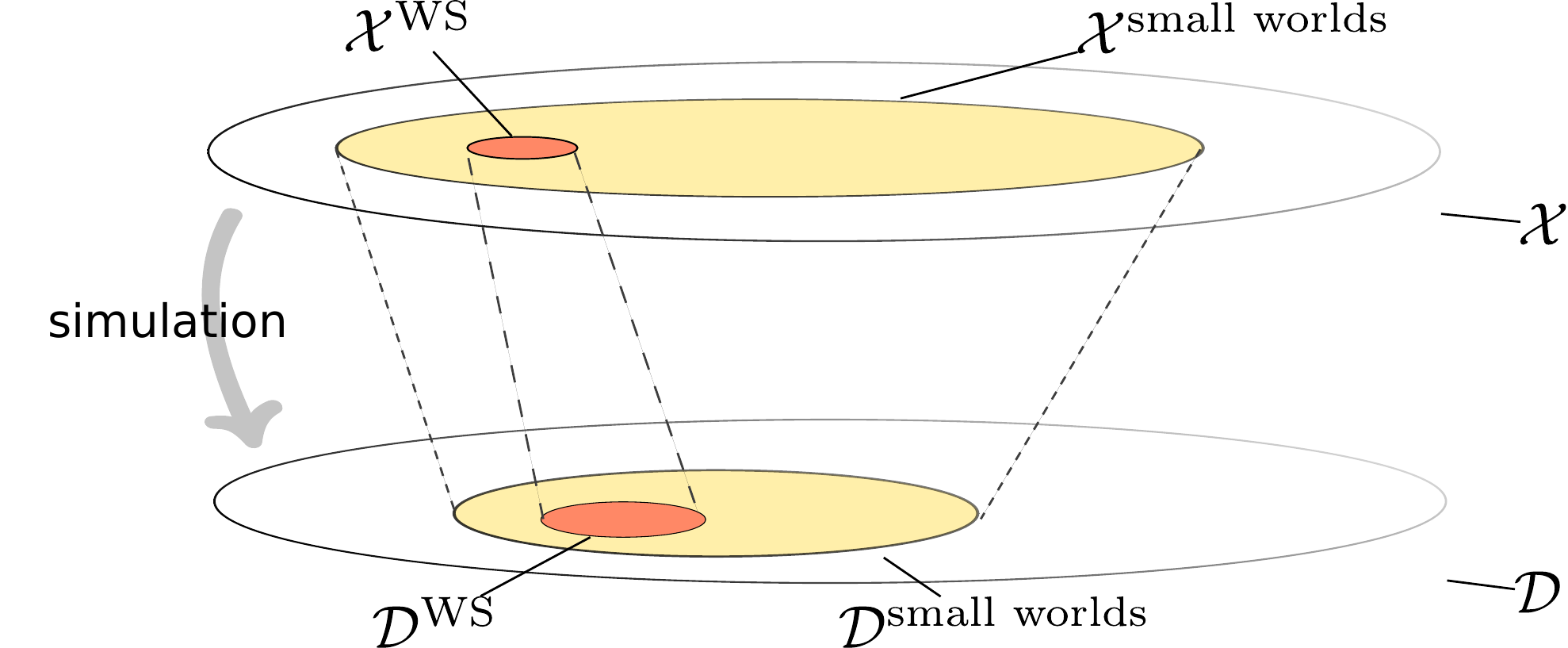}
}
\caption{Simulation of a model viewed as a projection from the space of networks $\mathcal{X}$ to the space of dynamics $\mathcal{D}$, with $\mathcal{D}^{\text{small world}}$ the possible dynamics using all the plausible networks, and $\mathcal{D}^{\text{WS}}$ the dynamics over networks $\mathcal{X}^{\text{WS}}$ generated by the Watts-Strogatz algorithm. The generator is said not representative of the class of networks if the dynamics $\mathcal{D}^{\text{WS}}$ does not covers the space of possible dynamics $\mathcal{D}^{\text{small world}}$ over all small-world networks.}
\label{fig:projectionD}
\end{figure} 

\subsubsection{Conclusivity of simulation results}

%\paragraph{}
Simulations of a model $M$ may be viewed as a projection (in an intuitive rather than mathematical understanding) from the space of networks $\mathcal{X}$ to the space of possible dynamics $\mathcal{D}$ (see figure~\ref{fig:projectionD}). As said previously, \textit{a model that wouldn't restrict the space $\mathcal{D}$ wouldn't play its filter role}~\citep{samuel_thiriot:bib_sma_simulation:legay_1973_1}; in other words, a model that predicts that everything is possible is of few interest, because it does not prove that some hypothesis encoded in the model is true or false (in the case of an explicative model) nor reduces the space of possible future dynamics (in the case of a predictive model). \textit{We name here ``conclusiveness'' the ability of simulations over a space of networks to play this filter role}. While the choice of networks is mainly argued to be a matter of plausibility (as reflected by the classical sentence ``we used a small-world network because it complies with evidence [...]''), it should also reduce the space of networks enough for simulation results to be conclusive.

\begin{figure}[t]
\ifthenelsehtml{
\includegraphics{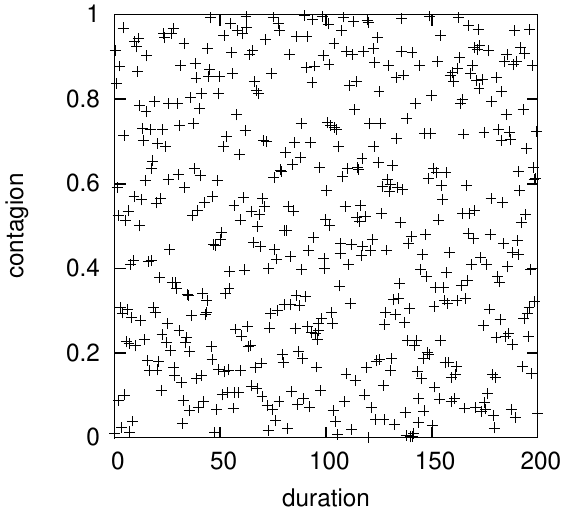}
\includegraphics{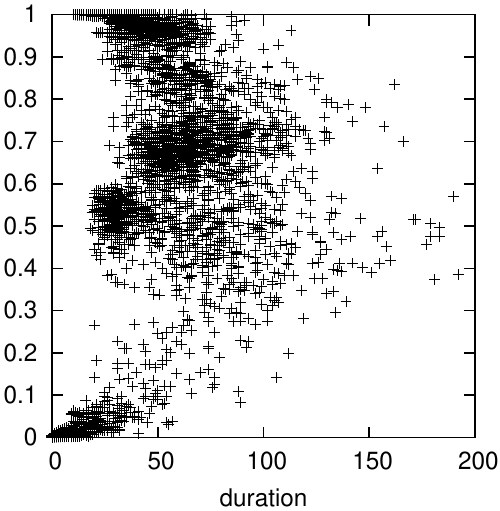}
\includegraphics{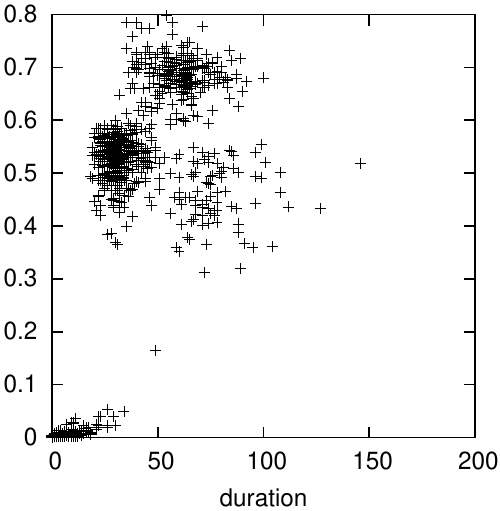}
} {
\includegraphics[height=\hautPrTroisCarres]{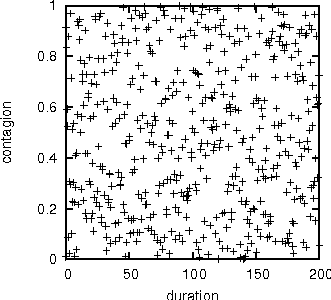}
\includegraphics[height=\hautPrTroisCarres]{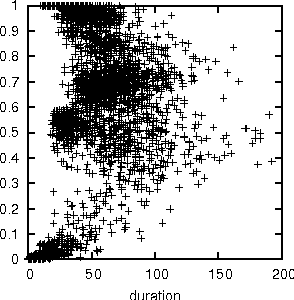}
\includegraphics[height=\hautPrTroisCarres]{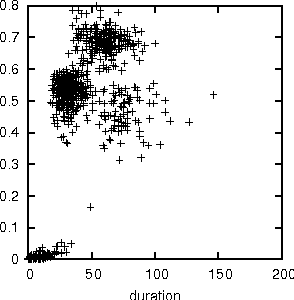}
}
\caption{Examples of simulation results in the the space of dynamics (duration, contagion) of an epidemic model. These examples illustrate the concept of simulation conclusiveness. \textit{(left)} an hypothetical case $\mathcal{D}_{\text{SIR}}^1$ where simulation results would be uniformly random; \textit{(center)} example of dynamics $\mathcal{D}_{\text{SIR}}^2$ over the space of small-worlds; \textit{(right)} dynamics $\mathcal{D}_{\text{SIR}}^3$ for one specific network generator.}
\label{fig:ex_sir}
\end{figure}

%\paragraph{}
To clarify this important point, let us take as an example an SIR model of epidemics. The dynamics of such a model may be studied in a two-dimensional space $\mathcal{D}_{\text{SIR}}$, the first dimension being the proportion of contaminated people (dead or recovered) at the end of the process and the second dimension the duration of the epidemics. Figure~\ref{fig:ex_sir} depicts, as an illustration, three examples of simulation results in this space $\mathcal{D}_{\text{SIR}}$. The left figure depicts a purely hypothetical case of uniform random results, learning us nothing about the possible dynamics in a population. The figure on the center is also of few interest, because it mainly means that the entire population may or not be contaminated, with variations in duration proportional to the extend of the epidemic. The right figure is more interesting, because it depicts a two-mode regime, with probable extension of the epidemics to about half of the population, and less probable extension to 80\% of the population. Obviously, the interpretation of the dynamics of a model is a difficult process that is a part of the modelling process; however, the gain in model usefullness (or simulations' conclusiviteness) between figures A and C is hardly contestable. 

%\paragraph{}
Let us now come back to the impact of the criteria for network selection $E^i$, by taking into consideration the fact that simulation results $\mathcal{D}_{\text{SIR}}^2$ are obtained by networks selected using permissive criteria $E^2$, while the more insightful results $\mathcal{D}_{\text{SIR}}^3$ are obtained with more precise criteria $E^3$. As we argued that $\mathcal{D}_{\text{SIR}}^2$ is useless while $\mathcal{D}_{\text{SIR}}^3$ is conclusive, then the criteria $E^2$ are also less interesting than $E^3$. In other words, \textit{our purpose as modelers is not only to define criteria $E^{\text{plausible}}$ that delimit a space of plausible networks, but $E^{\text{plausible}}$ should also be constrainsting enough for simulation results to be conclusive in the corresponding space of dynamics $\mathcal{D}_{\text{SIR}}^{\text{plausible}}$}.

% If the space of networks is too large, that is if the criteria of network plausibility $E$ is not precise enough, then the space $\mathcal{D}^{SIR}$ will contain short pandemics as well as slow and limited propagation of the illness. The purpose of the modeller is thus to determine a criteria for networks $E$ defining a space of networks $\mathcal{X}^{E} \subset \mathcal{X}$ that limits the space of dynamics $\mathcal{D}^E \subset \mathcal{D}$ to quite similar dynamics. 

\subsection{Experimental protocol\label{indoc:protocol_precise}}

%\paragraph{}
Having defined the qualitative process of generator choice in a more formal way, we can now clarify the problematics related to the interplay between criteria of network choice and the corresponding dynamics, and the experimental protocol that may answer these problematics. The experimental protocol is defined for given criteria of network choice $E^{\text{plausible}}$ that define a subspace of plausible networks $\mathcal{X}^{\text{plausible}} \subset \mathcal{X}$ and a model $M$ whom dynamics are studied in a space $\mathcal{D}_{\text{M}}$. The protocol to study the the three problematics exposed in a discursive way in introduction~(\pageref{indoc:problematics_intro}) is:
\begin{itemize}
\item Assessing the \textit{representativity of a generator} $G^j$ to a class of networks $\mathcal{X}^{\text{plausible}}$ defined by criteria $E^{\text{plausible}}$ comes to study by simulation whether the dynamics $\mathcal{D}_{\text{M}}^j$ over the networks $\mathcal{X}^{j}$ generated by $G^j$ are similar to the dynamics $\mathcal{D}_{\text{M}}^{\text{plausible}}$ supported by networks $\mathcal{X}^{\text{plausible}}$. The non-representativity of the generator $G^j$ will be proved if differences appear between $\mathcal{D}_{\text{M}}^j$ and $\mathcal{D}_{\text{M}}^{\text{plausible}}$ . Even if the space of plausible networks $\mathcal{X}^{\text{plausible}}$ cannot be extensively explored, it can be ``sampled'' by using various random network generators compliant with criteria $E^{\text{plausible}}$, each being tested with various parameter settings.

% Some examples of simulations over $\mathcal{X}^{small worlds}$ could be sufficient to proove a difference in dynamics. 
\item The \textit{inconclusiveness problem for a class of networks} may be highlighted by simply running the very same model on various networks in the space $\mathcal{X}^{\text{plausible}}$. One more time, the exploration $\mathcal{X}^{\text{plausible}}$ involves the use of several network generators, in order to sample the space with networks having different properties (in practice, one can use the samples created for the next step). If these simulations don't lead to conclusive results, the criteria for network choice $E^{\text{plausible}}$ would be proved not to lead to conclusive results, suggesting that the criteria for network choice should be refined.

\item The \textit{criteria refinability potential} can be assessed by defining more precise definition of the criteria $E^{\text{plausible}}$ and styding whether they lead, or not, to more conclusive results. For such a study to make sense, several precise criteria $E^{i}$ should be defined.  Each set of criteria defines a more precise subset of networks $\mathcal{X}^i \subset \mathcal{X}^{\text{plausible}} \subset \mathcal{X}$. As before, several random network generators should be configured for each $\mathcal{X}^i$. If simulations over these spaces $\mathcal{X}^i$ lead to coherent results, it would suggest that field work for measuring the quantitative values for criteria would improve simulation results. Else the original criteria would be proved to be insufficient, even when refined, for obtaining conclusive results, proving the urge to propose novel criteria for network choice. 
% 
% as specific combinaison of new indicators may hopefully lead to a better discrimination in the space of networks. Nevertheless, we can proove that a qualitative definition of the statistical indicators involved in the definition of the small-world phenomenon (namely clustering, average path length and density) does not leads to coherent results. Experiments may rely on simulations over subspaces $\mathcal{X}^i \subset \mathcal{X}^{small worlds}$ of small-world networks sharing precise characateristics $E^i$ lead to different results, and the assessment of the stability of dynamics $\mathcal{D}^i$ supported by each set of networks $\mathcal{X}^i$.
\end{itemize}

\section{Experimental settings for small-world\label{indoc:protocol_WS}}

\subsection{Application to small-worlds}

%\paragraph{}
We now apply the experimental protocol proposed before to the class of small-world networks. We first (\ref{indoc:netgens}) list the five network generators involved in the exploration of the space of small-world networks, then (\ref{indoc:spacesexploredWS}) detail the properties of the spaces of small-world which will be actually explored during the experiments. As shown by the various examples listed before (\ref{indoc:relatedwork}), agent-based modelling often use Watts-Strogatz\footnote{The seminal Nature paper that describes the Watts-Strogatz generator counts in the most cited papers in JASSS \citep{samuel_thiriot:bib_sma_simulation:meyer_2009_1}.\label{indoc:footnote_meyer}} and/or Barab\'asi-Albert networks to study the dynamics of models. In the case of small-world networks, the general problematics formalized before become:
\begin{itemize}
\item We argue of the use of Watts-Strogatz networks because of their plausibility; however, are simulations over WS networks representative to the space of small-world networks that we assume to be plausible ?
\item The real criteria for network choice involved here is the small-world phenomenon, Watts-Strogatz networks being only examples of these networks. If WS networks are not representative of the dynamics, are the dynamics over the space of small-world networks conclusive, or as conclusive as results obtained using WS networks~?
\item If these results are not satisfying, could they be improved by more precise criteria for network choice, like using precise combinations of network size, clustering rate and/or average path length~?
\end{itemize}

\subsection{Network generators\label{indoc:netgens}}

\begin{figure}[th]
\centering
\ifthenelsehtml{
\begin{tabular}{ccc}
\includegraphics{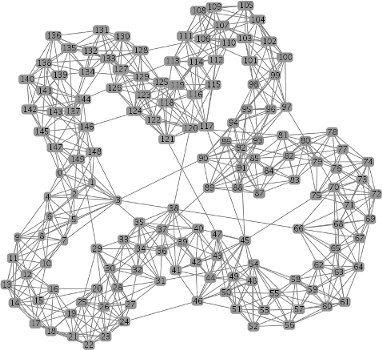} &
\includegraphics{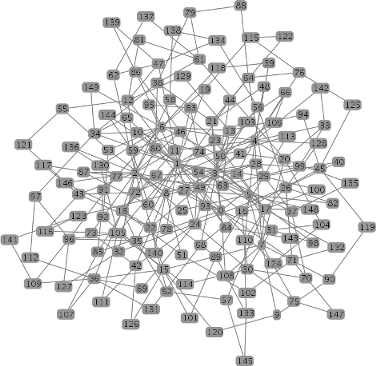} &
\includegraphics{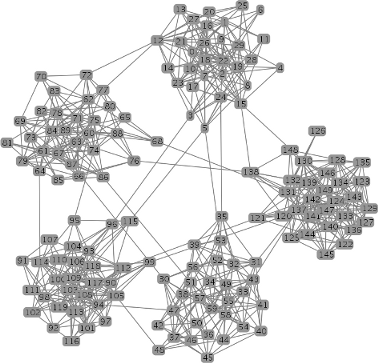} \\ 
WS: Watts-Strogatz &
BA: Barab\'asi-Albert &
SII: Simple Interconnected Islands\\
\end{tabular}
\begin{tabular}{cc}
\includegraphics{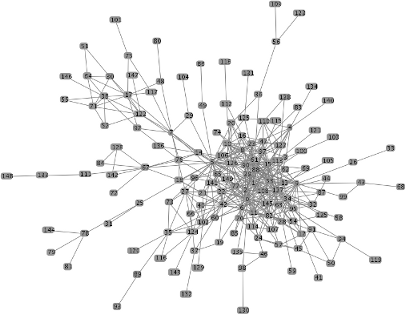} &
\includegraphics{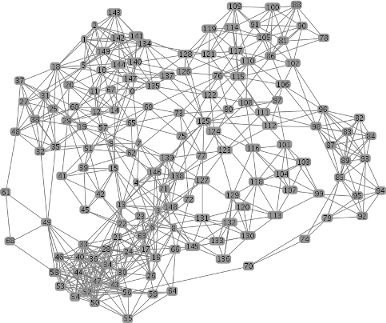} \\
FF: Forest Fire&
GRG: Geometric Random Graph\\
\end{tabular}
} {
\begin{tabular*}{\textwidth}{p{0.3\textwidth}p{0.3\textwidth}p{0.3\textwidth}}
\includegraphics[width=\largPrExReseauxA]{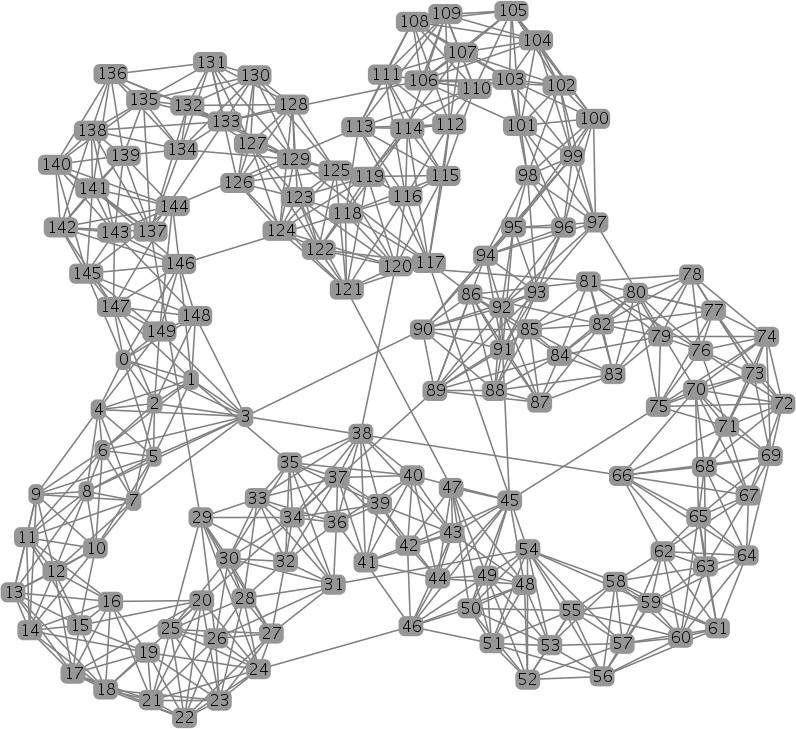} &
\includegraphics[width=\largPrExReseauxA]{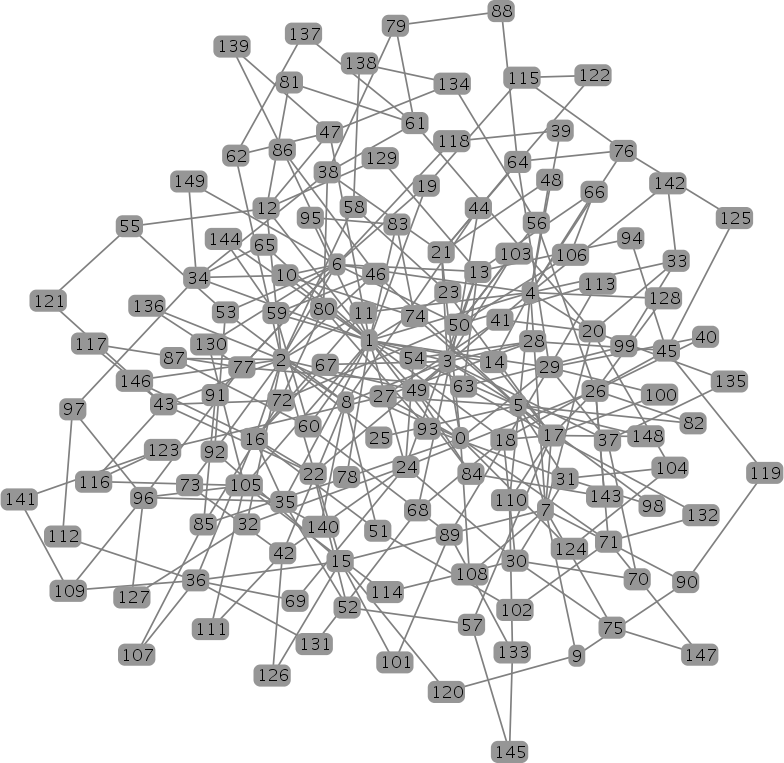} &
\includegraphics[width=\largPrExReseauxA]{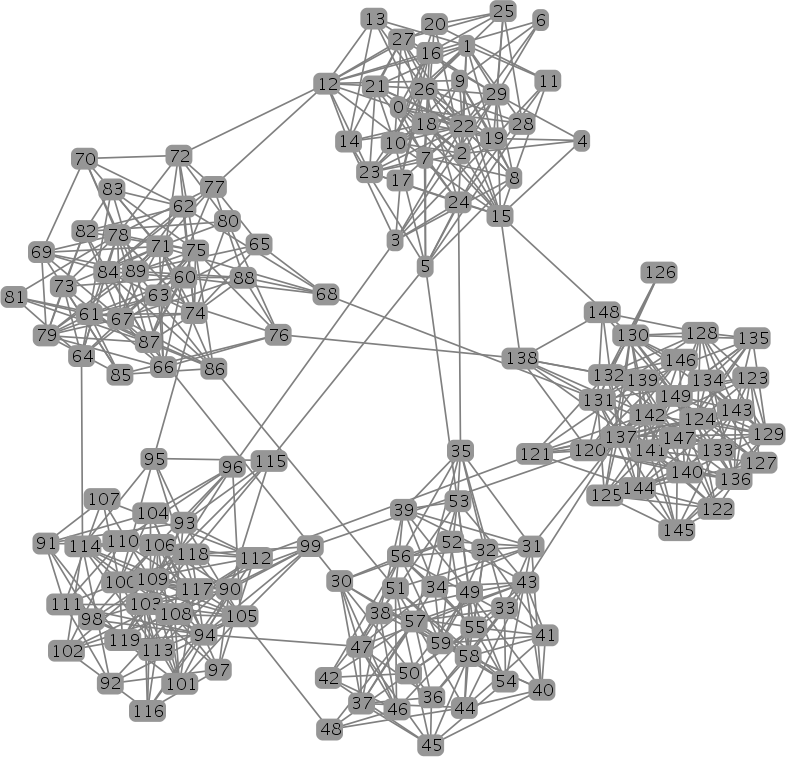} \\ 
WS: Watts-Strogatz &
BA: Barab\'asi-Albert &
SII: Simple Interconnected Islands\\
\end{tabular*}
\begin{tabular*}{\textwidth}{p{0.45\textwidth}p{0.45\textwidth}}
\includegraphics[width=\largPrExReseauxB]{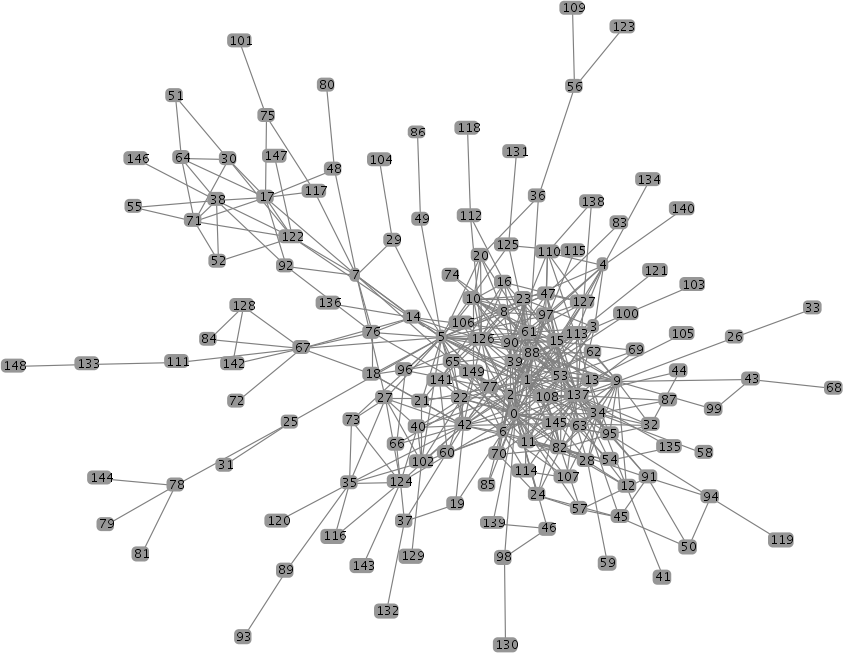} &
\includegraphics[width=\largPrExReseauxC]{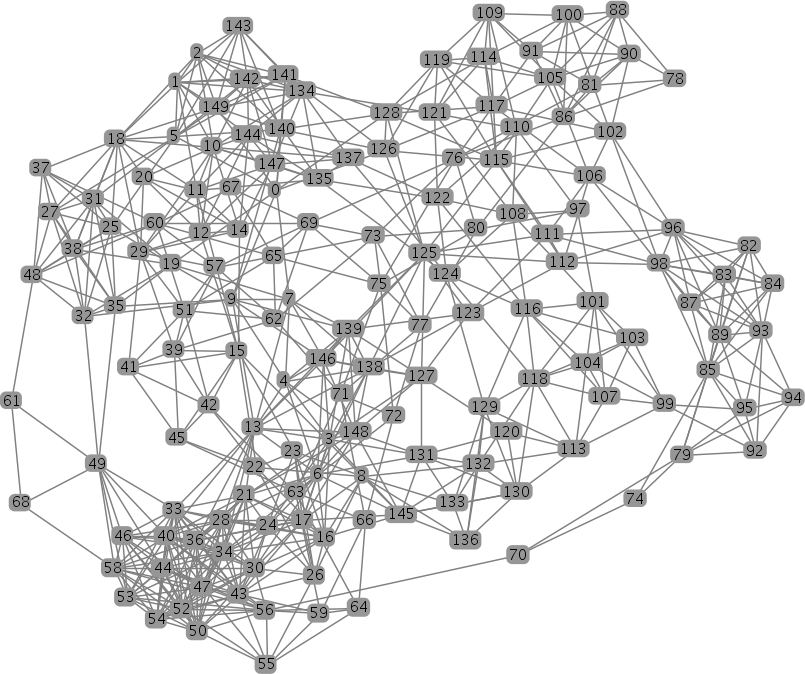} \\
FF: Forest Fire&
GRG: Geometric Random Graph\\
\end{tabular*}
}
\caption{Examples of small-world networks created by each generator used in this paper.}\label{fig:networks}
\end{figure}

%\paragraph{}
% There is no way to explore the whole space of small-world networks $\mathcal{X}^{small worlds}$ in an extensive way. We rather use different network generators that create networks having different characteristics (like the presence of communities or skewed distribution of degree), and parameter them to sample this space. We select five random network generators for sampling the spaces of networks. 
We use five network generators to sample the space of small-world networks. They are selected because of their wide use and the different properties of the network they generate (see examples in figure~\ref{fig:networks}), as well as for the flexibility they offer for tuning the characteristics of generated networks throught adequate parametering. 
% The two first, the Watts-Strogatz and Barab\'asi-Albert generators, are the most used in agent-based simulation. Geometric Random Graphs and the Forest Fire algorithm are less famous; the first generate spatialized networks with communities, while the second creates networks having at the same time a skewed distribution of degree, as core-periphery network, communities and assortativity of degree. We add a fifth generator of our own, the Simple Interconnect Island algorithm, that generate networks composed by interconnected communities. We provide here a quiet view of these generators and the typical properties of the networks they generate. Typical examples of the networks generated by these algorithms are depicted in figure~TODO. Note that, as classicaly done is social simulation, \textit{networks will be used in an undirected way}, and will only be used with relatively small sizes (less than thousands nodes).

\subsubsection{WS: Classical small-world networks}

%\paragraph{}
The famous algorithm proposed by Watts and Strogatz is likely the most used in social simulation\footref{indoc:footnote_meyer}. We use here the $\beta$-model \citep{samuel_thiriot:bib_sma_simulation:watts_1998_1,samuel_thiriot:bib_sma_simulation:watts_1999_2}, that requires as parameters $N$ the size of the network, $nei$ the neighborhood of the original lattice and $p^{rewire}$ the rewiring probability. This algorithm starts with a regular lattice of $N$ nodes in which nodes are connected with their $nei$ neighboors (thus having $2.nei$ degree). It then rewires each link with probability $p^{rewire}$. In case of rewiring, the link is disconnected from its target node and reconnected with any other node with uniform probability. The Watts-Strogatz generator was built to create networks having a short average path length and an high clustering rate. The distribution of degrees in large WS networks follows a Poisson law (see~\cite[p.~23]{samuel_thiriot:bib_sma_simulation:albert_2002_1} for an overview of Watts-Strogatz networks' properties). 

\subsubsection{BA: Barab\'asi-Albert scale-free networks}

%\paragraph{}
Scale-free networks, as defined by Barab\'asi and Albert, progressively replace or complete Watts-Strogatz networks in agent-based simulation. This generator complies with the scale-free (more exactly, fat-tailed or skewed) distribution of degree observed in several social networks (e.g. emails \citep{samuel_thiriot:bib_sma_simulation:ebel_2002_1} or sexual contacts~\citep{samuel_thiriot:bib_sma_simulation:liljeros_2001_1}). Their simple algorithm \citep{samuel_thiriot:bib_sma_simulation:barabasi_1999_1} implements one plausible explanation (among other explanations \citep{samuel_thiriot:bib_sma_simulation:keller_2005_1}) of this fact, by growing step by step the network, each novel node being connected with $m$ old nodes with a preferential attachment to nodes having already an high connectivity. The parameters of the models are $N$ the size of the network, $m$ the number of links added for each novel node and $\alpha$ the power of the preferential attachment (1 being linear). BA networks have a short average path length~\cite[p.~30]{samuel_thiriot:bib_sma_simulation:albert_2002_1}. They were said to be small-world networks because their clustering coefficient is higher than in a random graphs~\cite[p.~31]{samuel_thiriot:bib_sma_simulation:albert_2002_1}; however, the BA generator creates networks having a clustering much smaller than WS networks, for same size and density \citep{samuel_thiriot:bib_sma_simulation:kawachi_2004_1}. 
% Klemm \citep{samuel_thiriot:bib_sma_simulation:klemm_2002_1}

\subsubsection{GRG: Geometric Random Graphs}

%\paragraph{}
Geometric Random Graphs \citep{samuel_thiriot:bib_sma_simulation:penrose_2003_1} (GRG) are generated by dropping randomly $N$ points on a unit torus $[0,1]^2$, then connecting together all the points separated by less than a given threshold $radius$. GRG networks may be viewed as a metaphor for spatialized networks~\citep{samuel_thiriot:bib_sma_simulation:costa_2007_1} like connections between houses or villages in rural areas. As depicted in figure~\ref{fig:networks}, GRG contain sparse areas as well as strongly and totally interconnected sets of nodes that may be assimilated to communities. The distribution of degree in such a networks may be approximated with a Poisson law for large networks~\citep{samuel_thiriot:bib_sma_simulation:dall_2002_1}. For small networks, geometric random graphs have a strong clustering rate compared to their density. 
% The average path length remains quiet low for small networks, but grows significantly when the size of the network increases while the density of the network remains low. We will thus only use this example for small sizes of the networks ($N < 1000$).

\subsubsection{FF: Forest Fire}

%\paragraph{}
The Forest Fire model was proposed by Leskovec \citep{samuel_thiriot:bib_sma_simulation:leskovec_2007_1} as an algorithm that creates networks having most of the properties observed in real networks, including communities, skewed distribution of degrees, a core-periphery structure. As in the BA algorithm, the network is grown step by step, each new node $A$ being attached to $m$ old nodes. Moreover, each time a new link is created between $A$ and $B$, $A$ explores the outgoing and incoming neighboors of $B$. $A$ create links with outgoing nodes of $B$ with a \textit{forward probability} $p$, and also creates links with incoming nodes of $B$ with probability $p.r$, with $r$ the \textit{backward burning ratio}. As this step is ran recursively, $A$ is said to ``burn'' all the possible links.

\subsubsection{SII: Simple Interconnected Islands}

%\paragraph{}
We also provide a simple model\footnote{This algorithm is highly similar to the one proposed by Newman and Girvan \citep{samuel_thiriot:bib_sma_simulation:girvan_2002_1} as a testbed for community detection, in which two parameters drive the probability of existence of links respectively intra and inter communities. Our model facilitate the present experiments due to the guaranteed connectedness of the network.} that creates networks composed of several communities (sets of nodes having a strong density). This model, later named SII for Simple Interconnected Islands, starts by creating $n$ islands of identical size $size$. Each island is a random graph in which links exist with probability $p.in$. Each island is connected with all the other islands with $n.inter$ links, each being created between two nodes randomly picked from each island. Density and transitivity in SSI networks can easily be tuned by varying the $n$, $size$ and $p.in$ parameters, while the average path length may be tuned with $n.inter$. Its distribution of degree is nearly a Poisson-like one (as each island is a random network). This average path length remains low, because all the islands are interconnected. The example of SSI network with this algorithm depicted in figure~\ref{fig:networks} corresponds to $n=5$ islands, $n.inter=2$ links inter-islands, $size=350$ and $p.in=0.3$.

\subsection{Explored spaces\label{indoc:spacesexploredWS}}

%\paragraph{}
Having defined our experimental protocol and selected various networks generators, we present now the criteria for network choice that define the spaces of networks $\mathcal{X}^i \subset \mathcal{X}$ that will be tested in our experiments as samples of small world networks.s

\subsubsection{Space $\mathcal{X}^{small worlds}$ of qualitative small-worlds}

%\paragraph{}
The most cited definition\footnote{Note that we do not reused the more recent and disputable redefinition of small-world networks which adds the fat-tailed distribution of degree to the mandatory criteria for a network to be small-world~\citep{samuel_thiriot:bib_sma_simulation:amaral_2000_1}} of small-world is also the more qualitative one: \textit{``small-world networks are characterized by a short average path length and a high clustering rate''} \citep{samuel_thiriot:bib_sma_simulation:watts_1998_1,samuel_thiriot:bib_sma_simulation:watts_1999_1}. Note that this definition implicitly includes the constraints of low density which characterizes real networks \citep{samuel_thiriot:bib_psycho:wasserman_1994_1} (i.e. the network is sparse, the average degree $d$ being far lower than the network size \citep{samuel_thiriot:bib_sma_simulation:watts_1999_1}).

%\paragraph{}
The \textit{average path length} $l$ is the average of the geodesic distance separating every pair of nodes in the network. The ``short'' average path length refers to surprisingly short number of steps (lower than 10 given measures on available networks and experiments) observed in very large networks. This ``small-world effect'' was popularized by the Milgrams' experiment \citep{samuel_thiriot:bib_psycho:milgram_1967_1,samuel_thiriot:bib_sma_simulation:kochen_1989_1}, reproduced by experimentation \citep{samuel_thiriot:bib_psycho:dodds_2003_2,samuel_thiriot:bib_sma_simulation:travers_1969_1} and measured in real networks (e.g.~\citep{samuel_thiriot:bib_sma_simulation:leskovec_2008_2}). Mathematically, the ``short'' adjective was formalized as $L$ having approximately the same value than for an equivalent random graph \cite{samuel_thiriot:bib_sma_simulation:watts_1998_1,samuel_thiriot:bib_sma_simulation:watts_1999_1}, which was shown to scale as the logarithm of the number of nodes \cite{samuel_thiriot:bib_sma_simulation:erdos_1959_1,samuel_thiriot:bib_sma_simulation:albert_2002_1}. 

%\paragraph{}
The \textit{clustering coefficient} $C$ of a network, also named transitivity or clustering rate, reflects the probability that two nodes are connected, given that they are both connected to a same third. The clustering coefficient $C$, which is the average fraction of pairs of neighbors of a node which are also neighbors of each other \citep{samuel_thiriot:bib_sma_simulation:watts_1998_1}. This clustering rate was shown to be high in social networks \citep{samuel_thiriot:bib_sma_simulation:newman_2003_1}. An ``high'' clustering rate is mainly understood as being higher than in random networks with same size and density~\citep{samuel_thiriot:bib_sma_simulation:watts_1998_1}, but was also formalized as $C \gg O(N^{-1})$ \citep{samuel_thiriot:bib_sma_simulation:newman_2000_2}. 

%\paragraph{}
The general criteria of small-world networks will be noted $E^{small worlds}$. Put in a formal way, $E^{small worlds}$ contains the propositions: $E^{small worlds} = \{small worldsext{low density}, \text{short average path length}, \text{high clustering} \}$, each of these propositions being understood given the definition cited before.

% As explained in the experimental protocol, we will sample the corresponding space of small-worlds $\mathcal{X}^{small worlds} \subset \mathcal{X}$ using networks compliant with the small-world criteria $E^{small worlds}$. There is no way to explore the whole space of small-world networks $\mathcal{X}^{small worlds}$ in an extensive way. We will rather rely on several ``samples'' of this space, that is subspaces of small-world networks, each having specific statistical properties and containing networks generated by different algorithms.

\subsubsection{Spaces $\mathcal{X}^i \subset \mathcal{X}^{small worlds}$ of similar small-worlds}

\begin{figure}[ht]
\ifthenelsehtml{
\begin{tabular*}{\textwidth}{c}
\includegraphics{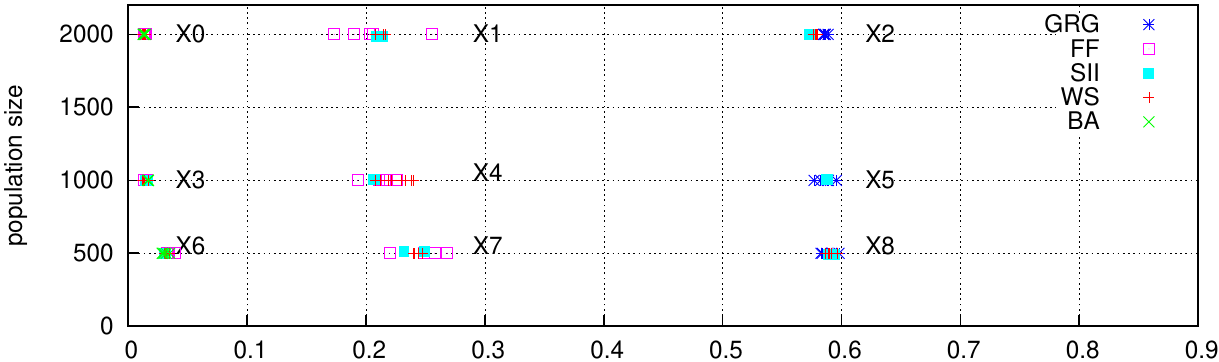}\\
\includegraphics{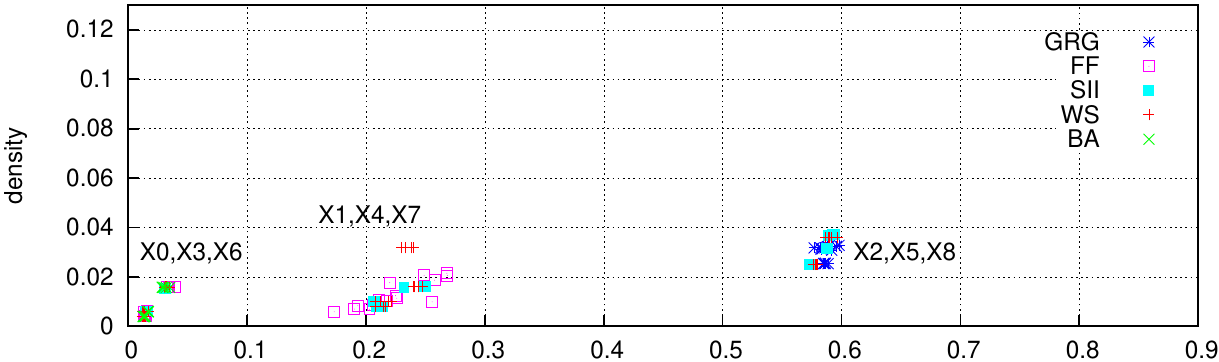}\\
\includegraphics{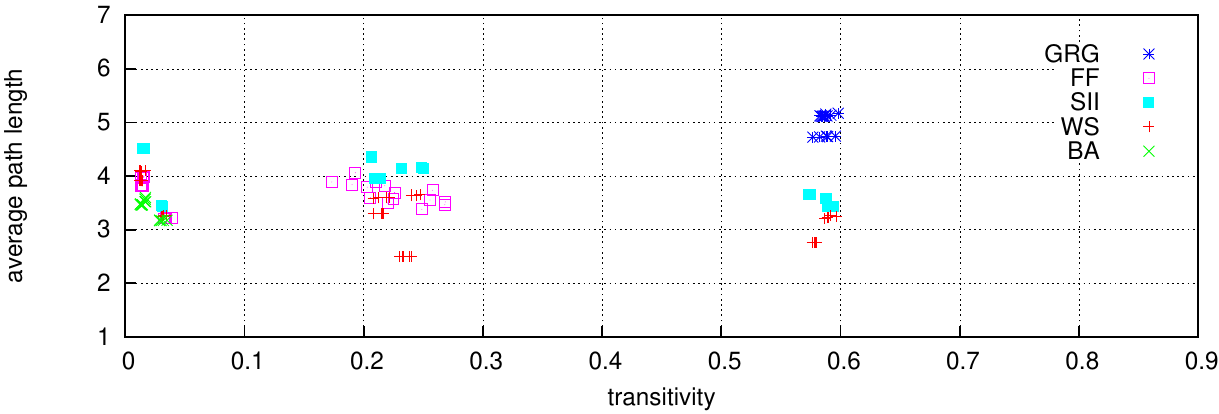}\\
\end{tabular*}
} {
\begin{tabular*}{\textwidth}{c}
\includegraphics[width=\textwidth]{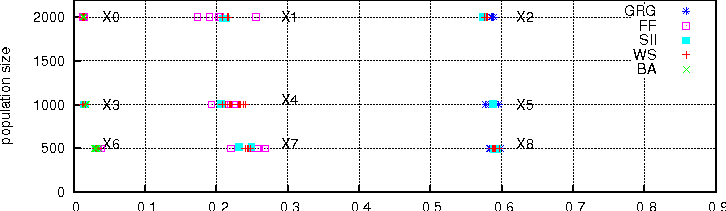}\\
\includegraphics[width=\textwidth]{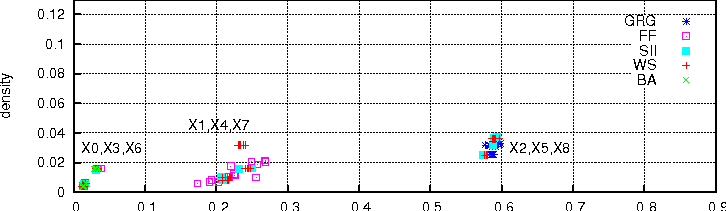}\\ 
\includegraphics[width=\textwidth]{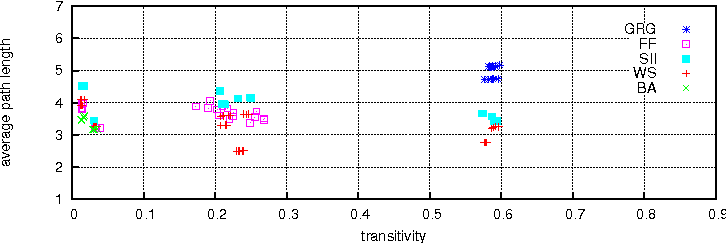}\\
\end{tabular*}
}
\caption{Properties of networks generated for all spaces $\mathcal{X}^i$ projected given their transitivity and \textit{(top)} size, \textit{(middle)} density and \textit{(bottom)} average path length.}
\label{fig:all_networks_density_trans_A}
\end{figure} 

%\paragraph{}
In order to compare then dynamics of more or less similar small-worlds, the various subspaces $\mathcal{X}^{i} \subset \mathcal{X}^{small worlds}$ of small-world networks are defined such that each space is characterized by the same values for network size transitivity as two other spaces. These various spaces are listed in table~\ref{tab:params_spaceA} along with the properties of networks they contain and the parameters used for generators. For each space, we configure all the possible network generators such the networks they generate are compliant with these properties. All the network generators cannot reach any part of this space. For instance, the Barab\'asi-Albert algorithm cannot generate networks having both a very high clustering rate such as 0.6 and a low density like 0.01; this generator is thus used only in spaces having a lower clustering rate. In a similar way, GRG can only generate networks having a very strong clustering rate, but cannot generate networks having a short average path length without having a very strong density. In the absence of predictive measures for all the generators \citep{samuel_thiriot:bib_sma_simulation:albert_2002_1}, the relevant parameters for each generator are tuned empirically. 

%\paragraph{}
The characteristics of the networks generated for each space are illustrated in figure~\ref{fig:all_networks_density_trans_A} given their clustering and size, density and average path length. In this figure, each point represents one example of network generated with the parameters listed in table~\ref{tab:params_spaceA}, 100 networks being generated per generator and subspace. Due to the stochastic component involved in these algorithms, each generator leads to more or less precise characteristics; these biases in the properties of the networks belonging to each subspace $\mathcal{X}^i$ are precised in table~\ref{tab:params_spaceA}.

% 
% %\paragraph{}
% We define the various spaces $\mathcal{X}^i \subset \mathcal{X}^{small worlds}$ of networks that share similar characteristics. These spaces are chosen such some of their properties are similar in order to enable the comparison of simulation results over different spaces. The size of networks has to be similar to enable comparison of dynamics, because the size of network is known to affect at least the duration of many social dynamics. 
% 
% % density / transitivity
% %\paragraph{}
% A same value of transitivity may be estimated more or less important given the density of the network: a complete network obvisouly has a transivity of 1, while a sparse one always has a low clustering rate (TODO citation ?). Every criteria in our list were thus built for networks having the same density and transitivity. 
% 
% % transitivity / size
% %\paragraph{}
% 
% % same density / transitivity / size
% %\paragraph{}

% We rather use different network generators that create networks having different characteristics (like the presence of communities or skewed distribution of degree), and parameter them to sample this space. Such an experimental protocol cannot be said to proove that the dynamics of a model is stable over the space of small-world networks, as this model could exhibit a different behaviour on a subspace of small-worlds that isn't covered by our samples. However, these samples are sufficient to proove that dynamics of one model are not similar over the space of small-world networks. 

\begin{table}
\footnotesize
\begin{tabular*}{\textwidth}{@{\extracolsep{\fill}}|cc|c c c c|}
\hline
\multicolumn{2}{|l|}{spaces $\mathcal{X}^i$} & 
\multicolumn{4}{c|}{criteria $E^i$ \& generator parameters} \\
\multicolumn{2}{|r|}{generators} &
$N$ &
$C$ &
$l$ &
$d$ \\
\hline
% $\mathcal{X}^0$ & 
% \multicolumn{6}{l|}{defined by $E^0$:$2000\pm0$, $0.015\pm0.004$, $3.7\pm0.28$, $0.004\pm0.00004$ } \\
% \hline
$\mathcal{X}^0$ & 
&
$2000\pm0$ & $0.015\pm0.004$ & $3.7\pm0.28$ & $0.004\pm0.00004$ \\
\hline
 &
BA &
\multicolumn{4}{l|}{$n=2000$, $p=1.0$, $m=4$} \\
&
WS & 
\multicolumn{4}{l|}{$n=2000$, $nei=4$, $p=0.5$} \\
&
FF & 
\multicolumn{4}{l|}{$n=2000$, $fw.prob=0.007$, $bw.factor=1$, $ambs=4$} \\
&
SII &
\multicolumn{4}{l|}{$n=4$, $M=500$, $p.in=0.016$, $n.inter=10$} \\

\hline 

$\mathcal{X}^1$ & 
&
$2000\pm16$ & $=0.21\pm0.032$ & $3.6\pm0.53$ & $0.008\pm0.003$ \\
\hline
&
WS & 
\multicolumn{4}{l|}{$n=2000$, $nei=8$, $p=0.181$} \\
&
FF & 
\multicolumn{4}{l|}{$n=2000$, $fw.prob=0.32$, $bw.factor=1$, $ambs=1$} \\
&
SSI &
\multicolumn{4}{l|}{$n=31$, $size=64$, $p.in=0.240$, $n.inter=2$}\\
\hline

$\mathcal{X}^2$ &
&
$2000\pm1$ & $0.58\pm0.01$ & $3.8\pm1.33$ & $0.026\pm0.0006$  \\
\hline
& GRG & 
\multicolumn{4}{l|}{$n=2000$, $radius=0.09$} \\
&
WS &
\multicolumn{4}{l|}{$n=2000$, $nei=25$, $p=0.040$} \\
&
SSI&
\multicolumn{4}{l|}{$n=23$, $size=87$, $p.in=0.581$, $n.inter=1$} \\

\hline
$\mathcal{X}^3$ &
&
$1000\pm1$ & $0.015\pm0.005$ & $4\pm0.58$ & $0.004\pm0.0022$ \\
\hline

&
BA &
\multicolumn{4}{l|}{$n=1000$, $p=1$, $m=3$} \\
&
WS & 
\multicolumn{4}{l|}{$n=1000$, $nei=3$, $p=0.495$}\\
&
FF & 
\multicolumn{4}{l|}{$n=1000$, $fw.prob=0.0036$, $bw.factor=1$, $ambs=3$}\\
&
SII &
\multicolumn{4}{l|}{$n=3$, $size=333$, $p.in=0.017$, $n.inter=50$}\\

\hline
$\mathcal{X}^4$ &
&
$1000\pm8$ & $0.22\pm0.03$ & $3.5\pm1.003$ & $0.016\pm0.017$ \\
\hline
&
SSI &
\multicolumn{4}{l|}{$n=24$, $size=42$, $p.in=0.235$, $n.inter=1$}\\
&
WS & 
\multicolumn{4}{l|}{$n=1000$, $nei=5$, $p=0.017$}\\
&
FF & 
\multicolumn{4}{l|}{$n=1000$, $fw.prob=0.37$, $bw.factor=1$, $ambs=1$}\\
\hline
 
$\mathcal{X}^5$ &
&
$1000\pm7$ & $0.58\pm0.013$ & $3.7\pm1.06$ & $0.004\pm0.03$  \\
\hline
&
GRG & 
\multicolumn{4}{l|}{$n=1000$, $radius=0.101$}\\
&
WS &
\multicolumn{4}{l|}{$n=1000$, $nei=16$, $p=0.038$}\\
&
SSI &
\multicolumn{4}{l|}{$n=19$, $size=53$, $p.in=0.6$, $n.inter=1$}\\
\hline

$\mathcal{X}^6$ &
&
$500\pm1$ & $0.03\pm0.011$ & $3.2\pm0.36$ & $0.015\pm0.001$ \\
\hline
&
BA &
\multicolumn{4}{l|}{$n=500$, $p=0.5$, $m=4$} \\
&
WS & 
\multicolumn{4}{l|}{$n=500$, $nei=4$, $p=0.420$}\\
&
FF & 
\multicolumn{4}{l|}{$n=500$, $fw.prob=0.01$, $bw.factor=0.5$, $ambs=4$} \\
&
SII &
\multicolumn{4}{l|}{$n=3$, $size=167$, $p.in=0.04$, $n.inter=80$}	\\

\hline
$\mathcal{X}^7$ & 
&
$500\pm13$ & $0.24\pm0.059$ & $3.8\pm0.6$ & $0.017\pm0.01$ \\
\hline
&
FF & 
\multicolumn{4}{l|}{$n=500$, $fw.prob=0.37$, $bw.factor=1.0$, $ambs=1$}\\
&
WS & 
\multicolumn{4}{l|}{$n=500$, $nei=4$, $p=0.15$}	\\
&
SII &
\multicolumn{4}{l|}{$n=19$, $size=27$, $p.in=0.29$, $n.inter=1$}\\
\hline

$\mathcal{X}^8$ & 
&
$500\pm7$ & $0.58\pm0.033$ & $3.9\pm1.29$ & $0.034\pm0.0043$ \\
\hline
&
WS & 
\multicolumn{4}{l|}{$n=500$, $nei=9$, $p=0.03$}\\
&
SSI &
\multicolumn{4}{l|}{$n=17$, $size=29$, $p.in=0.635$, $n.inter=1$}\\
&
GRG &
\multicolumn{4}{l|}{$n=500$, $radius=0.1$} \\
\hline 
\end{tabular*}
\caption{Characteristics $E^i$ of network spaces $\mathcal{X}^i$ used to explore the space of small-worlds $\mathcal{X}^{SW}$, with $N$ the number of nodes (network size), $C$ the clustering rate or transitivity, $l$ the average path length and $d$ the density.}
\label{tab:params_spaceA}
\end{table}

\subsection{Implementation}

%\paragraph{}
All the networks are generated and analyzed using the igraph package \citep{samuel_thiriot:bib_software:csardi_2006_1} for the statistical sofware named R \citep{samuel_thiriot:bib_software:Rteam_2009_1}. As most of random network generators in igraph create networks with redundant or self-links, those are removed during a simplification step. The simulations are driven with an ad-hoc simulation software interfaced with R for the generation of graphs. This software is developed in Java, and relies on various specialized libraries in order to improve reliability of results (e.g. random numbers are generated using the colt library). The implementation of each agent-based model was verified by comparing simulation with classical results from the literature.

\section{Experiments on small-worlds\label{indoc:experiments}}

%\paragraph{}
We present now successively the dynamics over these various spaces for three famous agent-based models, namely the \textit{epidemic model} (\ref{indoc:XP_SIR}), the \textit{opinion BC} (\ref{indoc:XP_OpinionBC}) and the \textit{Axelrod model of cultural dynamics} (\ref{indoc:XP_Axelrod}). For each model, we will define one unique space of parameters, which is intentionally selected for the model to be sensitive to the network of interaction (e.g. in the case of the epidemic model, setting the contagion parameter such as the entire population is for sure contaminated is of few interest). Indicators will be defined for each model to quantify the final state of the simulation. 
% Simulation results for each model are presented in a separated way, with an comprehensive viewpoint 

%\paragraph{}
Note that these experiments do not aim to explore the dynamics of the model itself; our ultimate purpose is to assess the relevance of the ``small-world'' criteria for network choice and the representativity of the Watts-Strogatz and Barab\'asi-Albert generators. Also, such an experimental protocol cannot be said to prove that the dynamics of a model is stable over the space of small-world networks, as this model could exhibit a different behavior on a subspace of small-worlds that is not covered by our samples. However, these samples remain sufficient to prove that dynamics of one model are not similar over the space of small-world networks.

\subsection{epidemic dynamics: SIR\label{indoc:XP_SIR}}

\subsubsection{Model}

%\paragraph{}
Epidemics are probably the most studied social phenomena in the stream of complex networks. We use a networked version of the simple and well-known SIR epidemic model \citep{samuel_thiriot:bib_sma_simulation:kermack_1927_1,samuel_thiriot:bib_sma_simulation:anderson_1991_1,samuel_thiriot:bib_sma_simulation:bailey_1957_1,samuel_thiriot:bib_sma_simulation:hethcote_2000_1}, in which each agent may be in one of the three states Susceptible (S), Infective (I) or Removed (R). All the agents but two are initialized in the Susceptible state, the last two being initialized in the Infective state. At each step of the simulation, all the links are activated in a random order, and the interaction is managemed between the agents connected by each link. If one of these agents is Infective and the other is Susceptible, then this last will shift to the infective state with probability $p^{contagion}=0.07$ at the end of the step. At the very end of the step, each infected agent may fall to the Removed state with probability $p^{removing}=0.2$. The simulation is stopped when the system becomes stable, that is when no interaction occurs during 100 steps. \newline
% enables to analyze the risk of a major outbreak, 

%\paragraph{}
Dynamics are studied against two criteria: the possible \textit{extend} of the pandemic and the \textit{duration} of this epidemic. Simulation results will then by studied in a two-dimensional space $\mathcal{D}^{SIR}$; results will be said to be conclusive if they reduce the space of possible dynamics in this space in either a quantitative (extend of the epidemic) or qualitative way (two mode regime - the epidemic remaining limited to a small fraction of the population or reaching a large part of it - or one mode). 
% 
% %\paragraph{}
% Many studies investigated epidemic dynamics over different networks \citep{samuel_thiriot:bib_sma_simulation:newman_2002_3} small-world  or scale-free networks. TODO. However, we are not interested in the behaviour of the model, but on the sensitivity and conclusivness (todo???) of the dynamics given the criteria of network choice.  

% Previous studies focused on the influence of the characteristics of networks on SIR dynamics, notably on assortativity (300), multiplex networks (21, 314). Among other, the epidemic model has not epidemic threshold when ran over most scale-free networks 
% TODO revoir neuman_2003_5 pour plus. 
% Dynamics in the population may range from no contagion to a total extend to the whole population. The duration of the epidemic may also change depending on the network state and the position of the first two infective agents. From a qualitative viewpoint, the behaviour of this model may exhibit a two-mode regime, with the epidemic remaining limited to a small number of agents (which is much more probable if the network contains communities) or extending to the larger part of the network. However, a one mode regim is also possible, with the contagion having a similar probability to extend to any propotion of the population. We study the dynamics of the SIR model on two main dimensions of interest: the proportion of the population which is reached at the end of simulation, and the duration of the epidemic (in steps). From a practical viewpoint, TODO.

\subsubsection{Results on the whole space of small-worlds $\mathcal{X}^{small worlds}$\label{indoc:sir_whole}}

\begin{figure}[ht]
\ifthenelsehtml{
\includegraphics{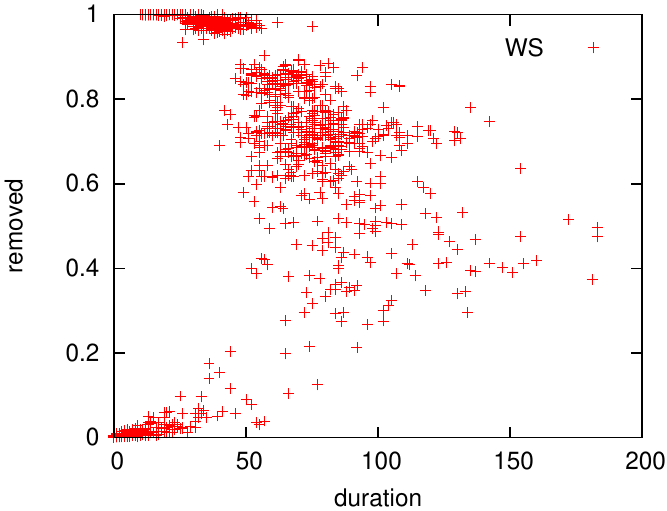} 
\includegraphics{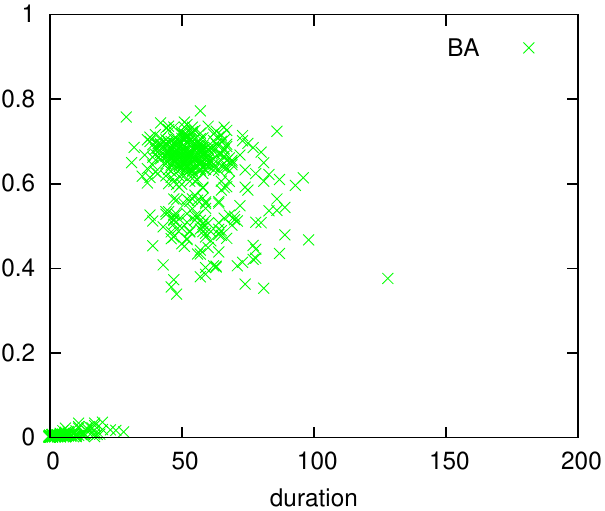}  
\includegraphics{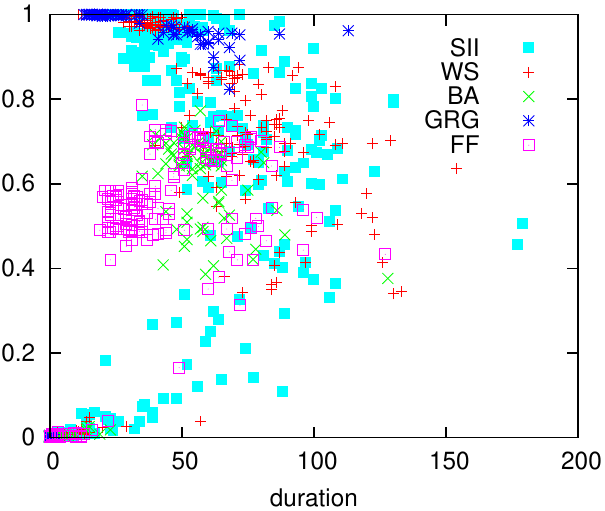}
} {
\includegraphics[height=\hautPrTrois]{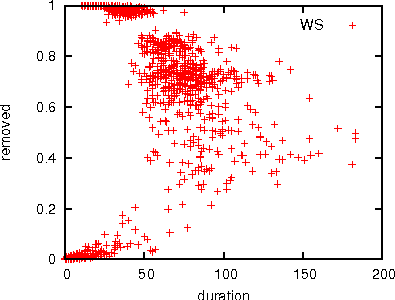} 
\includegraphics[height=\hautPrTrois]{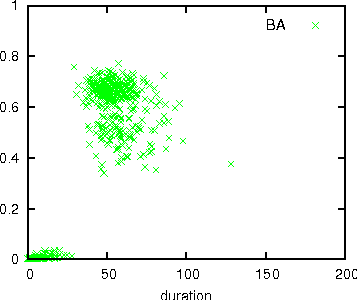}  
\includegraphics[height=\hautPrTrois]{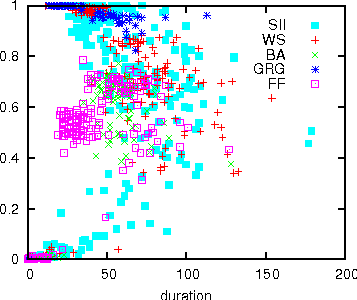}
}
\caption{Dynamics $\mathcal{D}_{\text{SIR}}$ of the SIR model supported by \textit{(left)} Watts-Strogatz networks, \textit{(center)} Barab\'asi-Albert networks and \textit{(right)} various small-world $\mathcal{X}^{qWS}$. Each dot in these figures represents the result of one simulation.}
\label{fig:SIR_spaceA_contagion_duration}
\end{figure} 

%\paragraph{}
Figure~\ref{fig:SIR_spaceA_contagion_duration} depicts, from left to right, simulation results obtained using Watts-Strogatz, Barab\'asi-Albert and all small-worlds. These results include all the parameter settings described in table~\ref{tab:params_spaceA}. As shown in the left figure, simulations over Watts-Strogatz networks suggest a two-mode regime, with an high probability for the pandemic to remain very limited (1\% of the population) or total (more than 95\% of the population). While the process in both these cases appear to be quiet quick ($< 50$ steps), some simulations using small networks suggest a third possibility, that is a diffusion ranging probably from 50\% up to 90\% of the population with an average duration of $~70$ steps. These last results are very similar with the simulation results using BA, which also predict a two-mode regime and possible extent ranging from 50\% to 80\% of the population. These simulation results, however, appear to be very specific when compared to the dynamics over other all small-worlds (right). With the same model and same parameters, much more intermediate states appear with more or less extension and more scattered durations. This comparison answers our first question in the case of SIR: \textit{epidemic dynamics supported by the BA and WS networks are not representative of dynamics in the larger space of small-worlds.}

%\paragraph{}
Simulation results on all sampled small-worlds (same figure~\ref{fig:SIR_spaceA_contagion_duration}, right) appear to be inconclusive: contagion ranges from 0\% to 99\% of the population. This duration is less than linear given the contaminated population because of the small-world effect, but varies as much as from 20 to 120 steps. In other words, \textit{the small-world phenomenon, in its qualitative understanding, does not constraints the space of networks enough to lead to conclusive results.}

\subsubsection{Results for small-worlds having the same characteristics $\mathcal{X}^i$}

\begin{figure}[ht]
\ifthenelsehtml{
\begin{tabular*}{\textwidth}{lcr}
\includegraphics{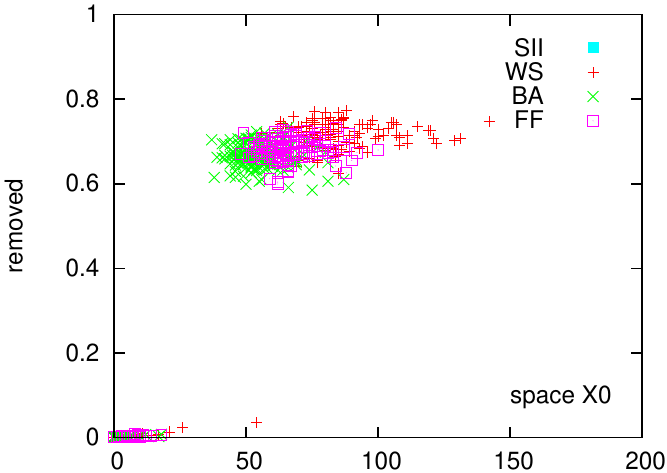}&
\includegraphics{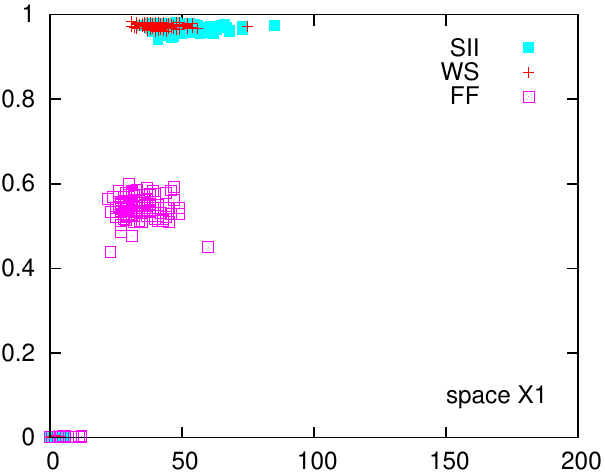} &
\includegraphics{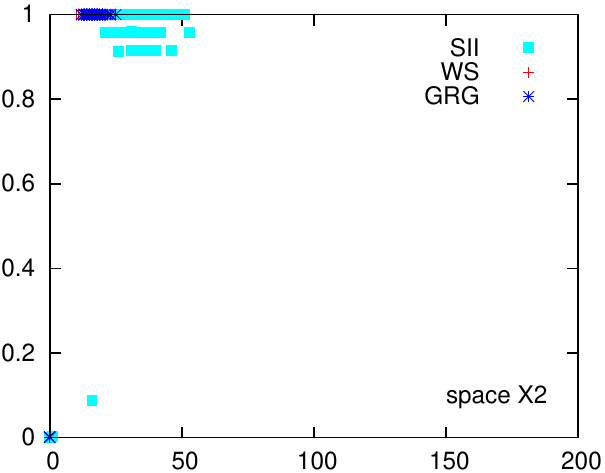} \\
\includegraphics{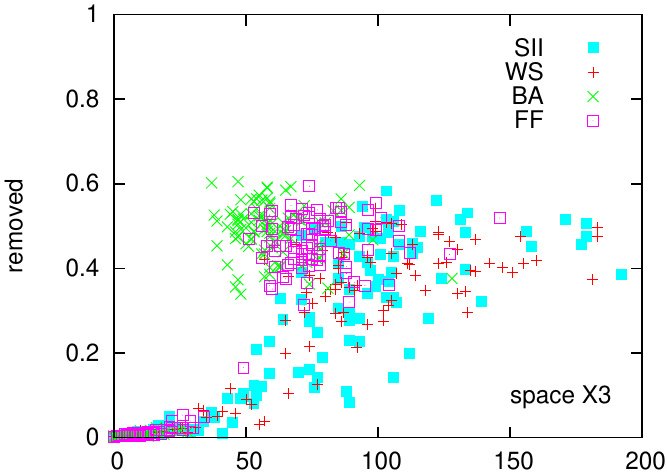} &
\includegraphics{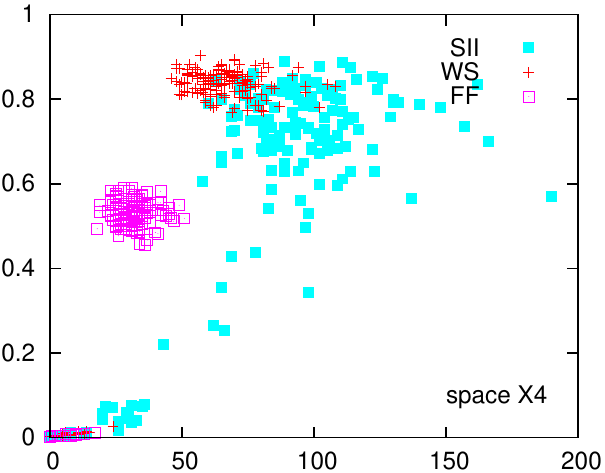} &
\includegraphics{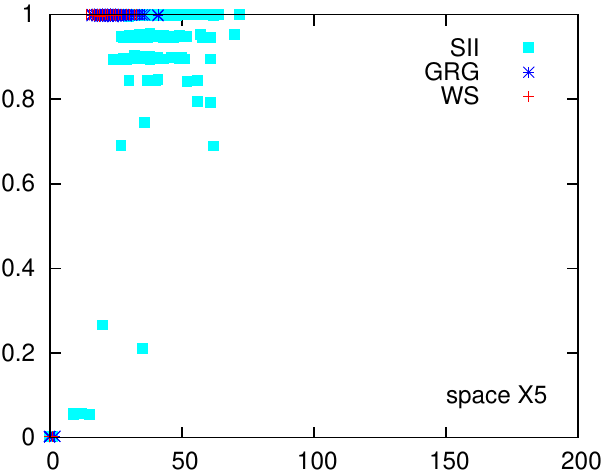} \\
\includegraphics{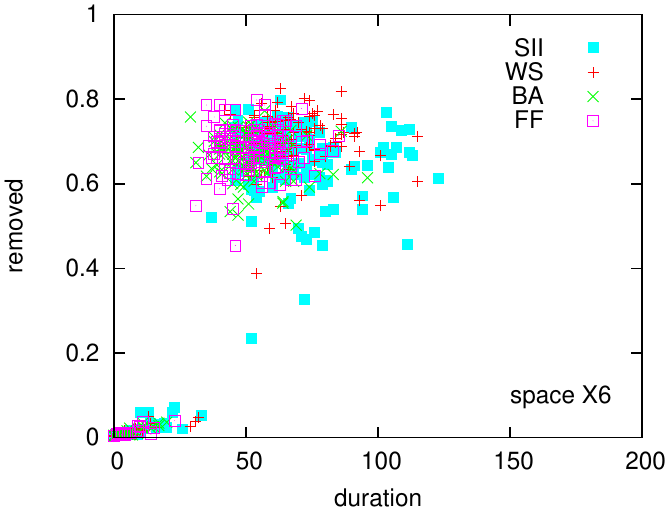} &
\includegraphics{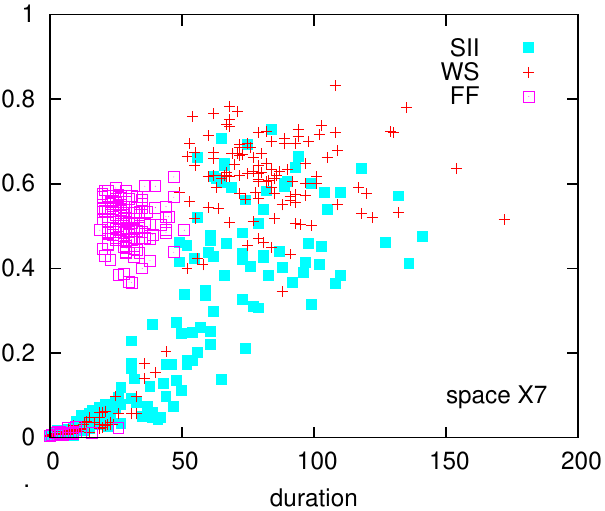} &
\includegraphics{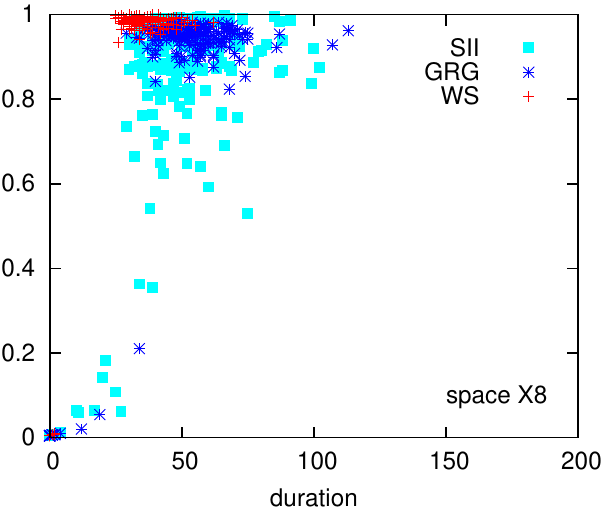} \\
\end{tabular*}
} {
\begin{tabular*}{\textwidth}{lcr}
\includegraphics[width=\mosaicHa]{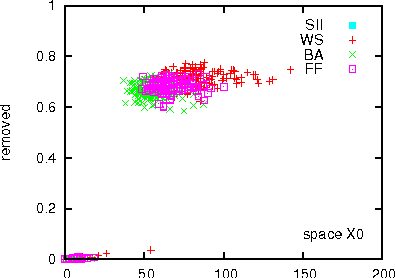}&
\includegraphics[width=\mosaicHb]{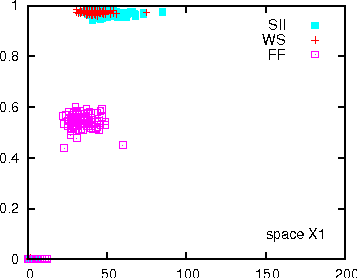} &
\includegraphics[width=\mosaicHc]{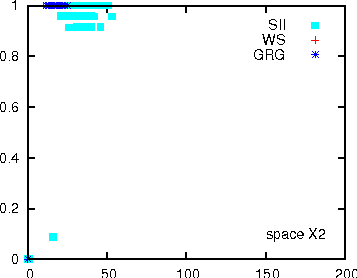} \\
\includegraphics[width=\mosaicMa]{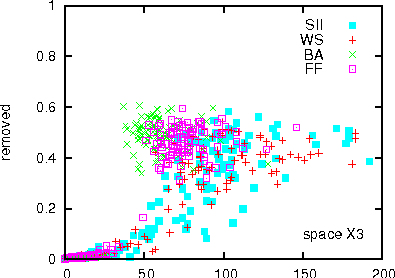} &
\includegraphics[width=\mosaicMb]{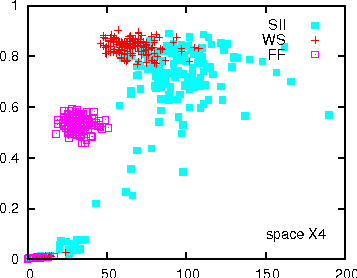} &
\includegraphics[width=\mosaicMc]{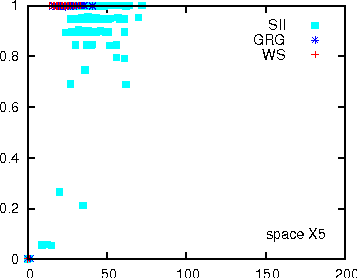} \\
\includegraphics[width=\mosaicBa]{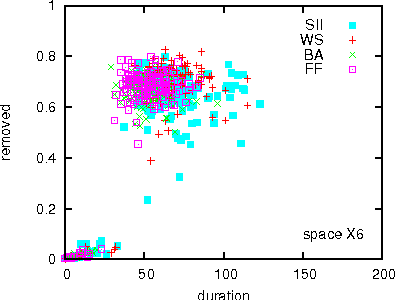} &
\includegraphics[width=\mosaicBb]{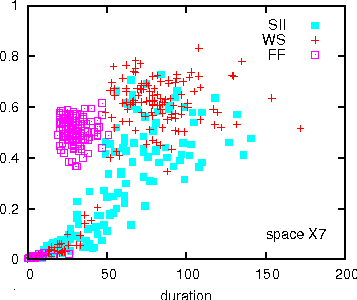} &
\includegraphics[width=\mosaicBc]{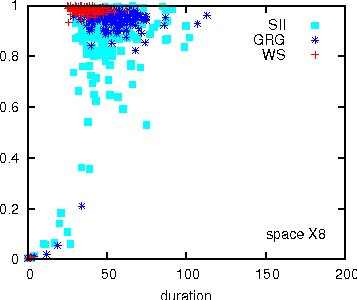} \\
\end{tabular*}
}
\caption{Dynamics $\mathcal{D}_{SIR}^{small worlds}$ of the SIR model given the size of networks. From top to bottom, spaces in each line correspond respectively to networks with 2000, 1000 and 500 nodes. From left to right, columns contain networks having the same clustering rates ($~0.01$, $~0.21$ and $~0.58$). Note that the results of WS networks are hidden by the GRG ones for spaces $\mathcal{X}^5$ and $\mathcal{X}^2$.}
\label{fig:SIR_spaceAx_contagion_duration}
\end{figure} 

%\paragraph{}
As depicted in figure~\ref{fig:SIR_spaceAx_contagion_duration}, spaces $\mathcal{X}^0$, $\mathcal{X}^3$ and $\mathcal{X}^6$ (left column) having a transitivity rate of  $~0.01$ all lead to qualitatively and quantitatively similar results whatever the network generator: bimodal regime, same extends, and even same high dispersion of results in the case of space $\mathcal{X}^3$. However, simulation results for spaces characterized by an high transitivity $~0.58$ ($\mathcal{X}^1$, $\mathcal{X}^4$, $\mathcal{X}^7$ in the center column) exhibit higher discrepancies. Dynamics over FF systematically lead to an extend of about half the others, because the propagation of the epidemic is unlikely in the periphery of these networks.  In spaces $\mathcal{X}^4$ and $\mathcal{X}^7$ which lead to bi-modal regimes, simple islands exhibit a one-mode regime, the extend of the simulation ranging from 0 to 0.8 with equal probability. This fact is due to the community-based structure which makes more probable the diffusion to all members of a community, but less probable the transmission to another community. Given these observations, simulations over networks of transitivity $~0.21$ are definitely inconclusive, as the model predicts an extend of 0\%, 50\% or all the population, with duration ranging from 30 to 150 steps. In the spaces depicted in the right column, which group the networks having a very high clustering rate, results are quiet coherent for all generators but the SII networks, because of the phenomenon observed in the center column.
% 
% \paragraph*{TODO}
% Note that studying results line by line enable to observe the coherency of results for networks of the very same size, but having different clustering rates. No

%\paragraph{}
As a conclusion, \textit{specific combinations of parameters may lead to stable results}, as for spaces $\mathcal{X}^0$, $\mathcal{X}^3$ and $\mathcal{X}^6$. This observation cannot be generalized, however: the examples of spaces $\mathcal{X}^1$, $\mathcal{X}^4$ and $\mathcal{X}^7$ prove that \textit{small-world networks having the same clustering rate, short average path length, density and size may lead to inconclusive results}. No absolute law may even be found to explain these differences; for instance, FF in the center column leads to a lower extend of the epidemics, which isn't the case in the left column. 

\subsection{Opinion BC\label{indoc:XP_OpinionBC}}

\subsubsection{Model}

%\paragraph{}
The Bounded Confidence model \citep{samuel_thiriot:bib_sma_simulation:deffuant_2000_2,samuel_thiriot:bib_sma_simulation:weisbuch_2002_1,samuel_thiriot:bib_sma_simulation:lorenz_2007_1} describes the dynamics of continuous opinions in a networked population. In this model, each agent holds a continuous opinion $x^i \in [0:1]$. The initial opinion are initialized with an uniform probability. At each step of the simulation, one link is randomly picked from the network, and the interaction between the connected agents is managed. If the opinions of the two agents $x$ and $x'$ are too different, that is if $|x-x'| \geq d$, then the agents don't influence each other. Else the opinion of each agent is changed in the direction of the other one: $x = x+\mu.(x'-x)$ and $x' = x'+\mu.(x-x')$. The dynamics of the Opinion BC model were extensively studied \cite[p.~80]{samuel_thiriot:bib_sma_simulation:boccaletti_2006_1}. The convergence parameter $\mu$ determines how quick the opinions of agents converge, and seems to have a negligible impact on the issue of the simulation, while the threshold $d$ directly determines the number of opinions. 
% : if $d > 0.5$, all opinions converge to a central one, with a number of opinions that survive various as $1/d$. \citep{samuel_thiriot:bib_sma_simulation:bennaim_2003_1}. For all the simulations, we set $\mu=0.4$ and $d=0.2$.

%\paragraph{}
We define two indicators for tracking simulations of the BC model: the \textit{number of major opinions} and the \textit{total number of opinions} (whatever the number of agents sharing them). The rational of these indicators is to investigate, not only the number of major opinions that may coexist in the same population, but also the propensity of a structured population to allow the survival of minor opinions. Minor opinions may survive in a population when agents are weakly connected to the core of the network, or simply because some agents ``hesitate'' between two major opinions, the simulation alternatively changing the opinion of these agents towards one opinion then the other. To measure the number of opinions, we discretize the space of opinions in 200 slices, and count the number of agents in each slice. This number of agents is then normalized by the maximum count of agents in a slice; major opinions are defined as opinions shared by at least one third of this proportion, while minor opinions are all slices that are not empty. Note that contiguous slices are considered to be one unique opinion. We stop the simulation after 100,000 steps; as this number is fixed, the existence of many opinions at stop may reflect simulations where the required time for convergence is higher.

\subsubsection{Dynamics for all networks}

\begin{figure}[ht]
\ifthenelsehtml{
\includegraphics{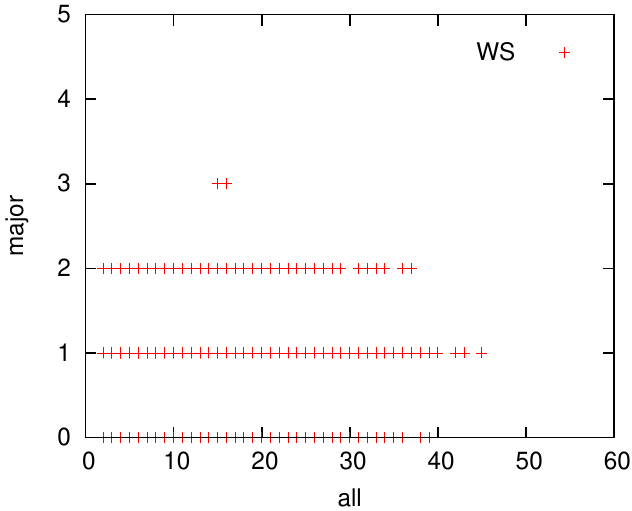} 
\includegraphics{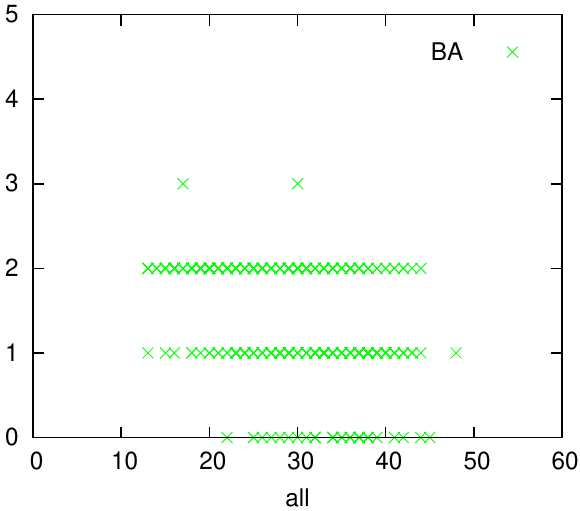} 
\includegraphics{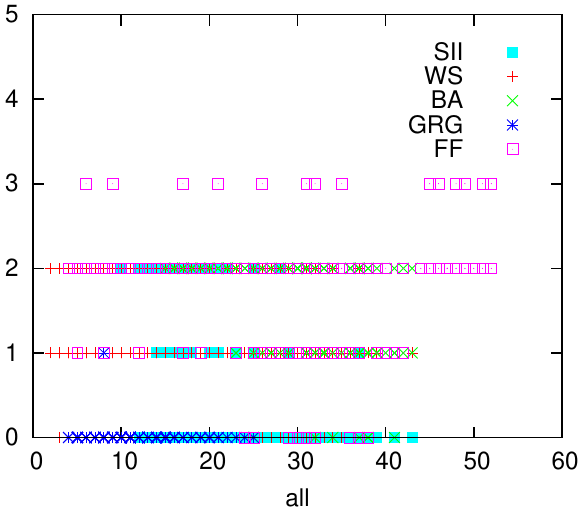} 
} {
\includegraphics[height=\hautPrTrois]{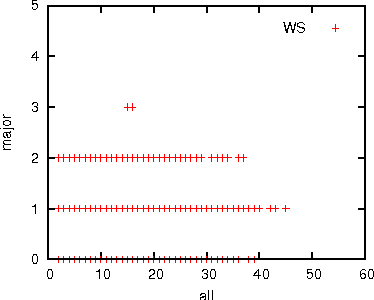} 
\includegraphics[height=\hautPrTrois]{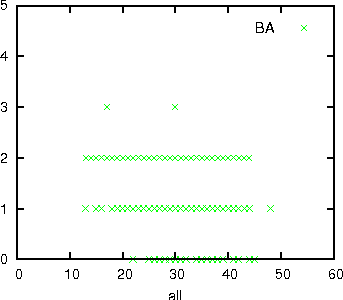} 
\includegraphics[height=\hautPrTrois]{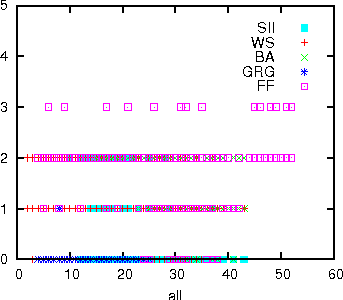} 
}
\caption{Dynamics $\mathcal{D}_{opinion}$ of the Opinion BC model supported by \textit{(left)} Watts-Strogatz networks, \textit{(center)} Barab\'asi-Albert networks and \textit{(right)} various small-world $\mathcal{X}^i$. Each point represents the result of one simulation.}
\label{fig:OpinionBC_spaceA}
\end{figure} 

%\paragraph{}
Simulation results for Watts-Strogatz and Barab\'asi-Albert networks, depicted in figure~\ref{fig:OpinionBC_spaceA} (left and center), are mainly different in the number of minor opinions (ranging from 0 to 45 for WS and 10 to 45 for BA), meaning that BA always maintain many minor opinions while WS have an higher probability to converge towards several opinions in the same duration. The average number of opinions is also different, WS leading with nearly equal probability to 0 (no major opinion), 1 or 2 major opinions, while BA has ~45\% chances to lead to 2 major opinions, ~40\% for 1 opinion and only ~14\% for no major opinion. These results for WS and BA are different from the dynamics observed for any network in our sample (figure~\ref{fig:OpinionBC_spaceA} right), in which the probability for simulations not to converge to major opinions is higher, as is the total number of opinions. Incidentally, only FF networks lead with significant probability to the coexistence of three major and many minor opinions. In these results, \textit{neither WS nor BA is representative of the results observed for all small-worlds}.

%\paragraph{}
All these results suggest that no more than three major opinions can survive in a structured population (in this experimental setting), the configuration with 3 major opinions being far less probable than for 0 to 2 opinions. They also learn us that no more than 50 minor opinions can survive in the population. More informative results are provided by the equiprobability of each number of opinions in the case of WS networks, or by the impossibility to have less than 12 minor opinions for BA networks. As results over all small-worlds do not provide such a discrimination in the space of dynamics, dynamics over this space may be argued to be less discriminative and conclusive than the other ones.

\subsubsection{Dynamics for similar small-worlds $\mathcal{X}^i$}

\begin{figure}[h]
\ifthenelsehtml{
\begin{tabular*}{\textwidth}{rrr}
\includegraphics{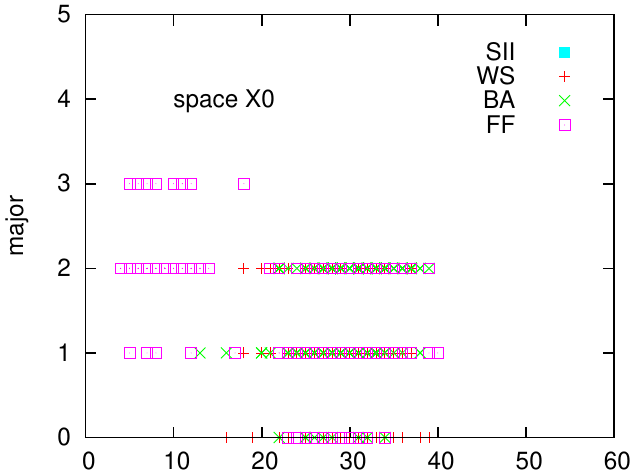} &
\includegraphics{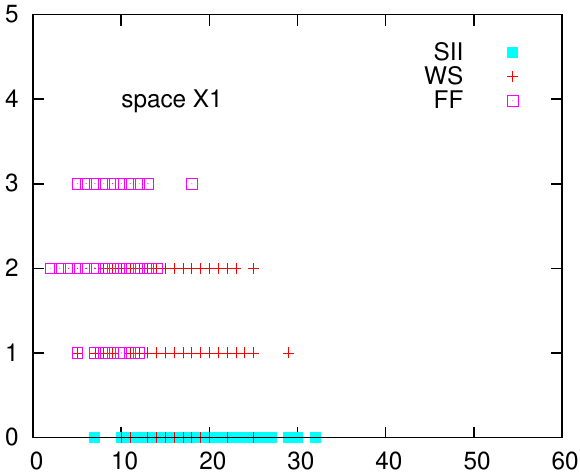} &
\includegraphics{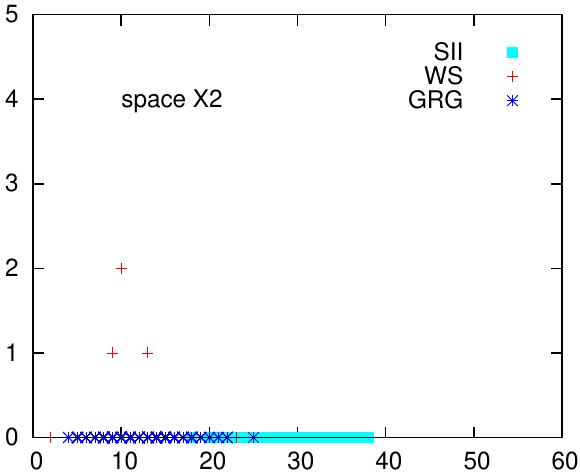} \\
\includegraphics{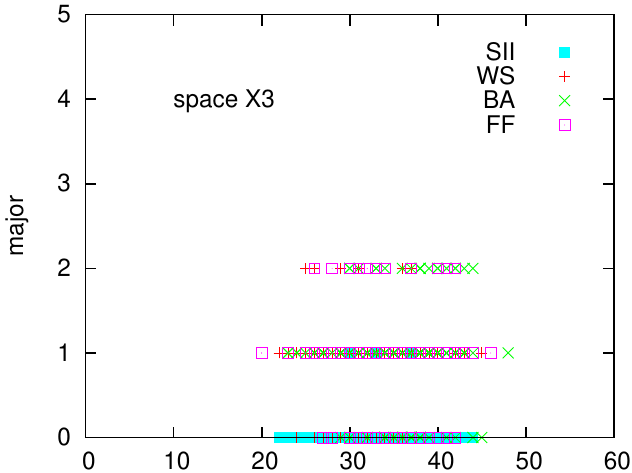} &
\includegraphics{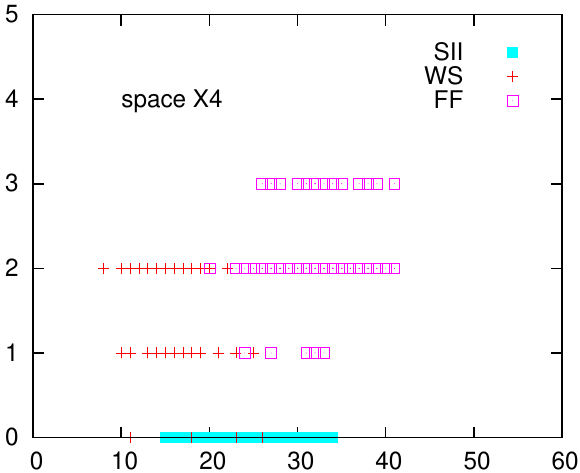} &
\includegraphics{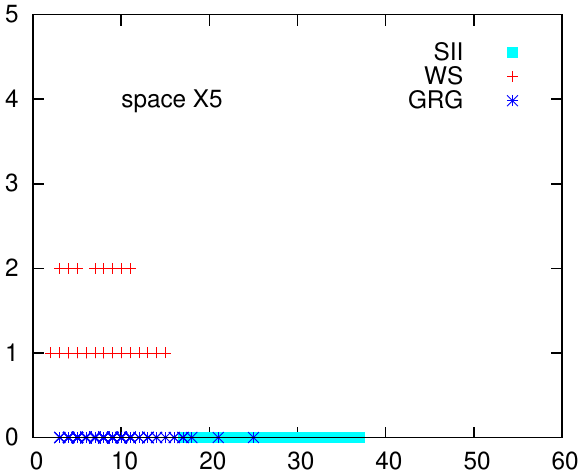} \\
\includegraphics{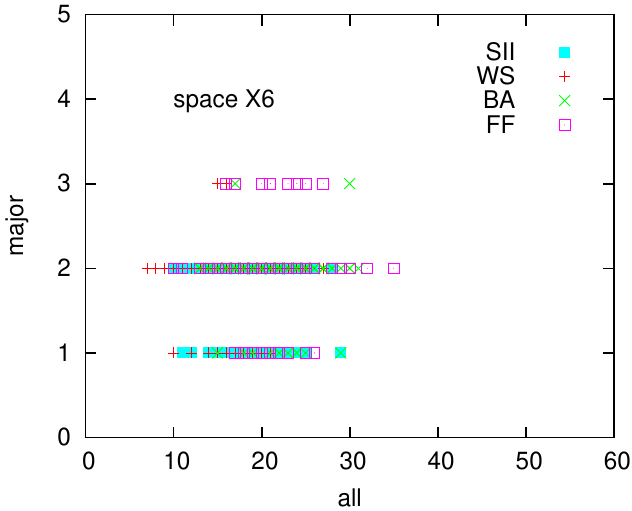} &
\includegraphics{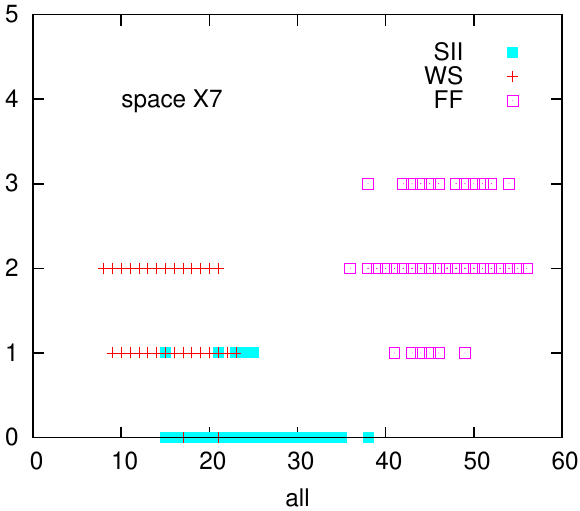} &
\includegraphics{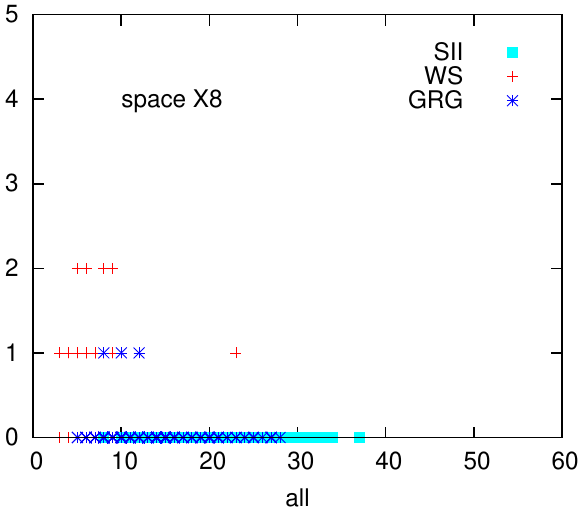} \\
\end{tabular*}
} {
\begin{tabular*}{\textwidth}{rrr}
\includegraphics[width=\mosaicHa]{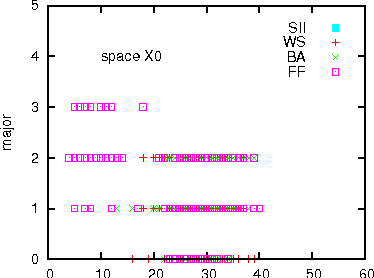} &
\includegraphics[width=\mosaicHb]{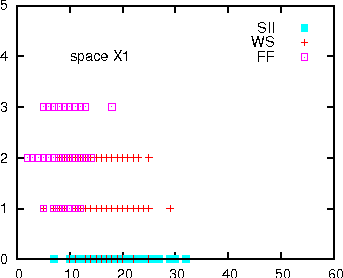} &
\includegraphics[width=\mosaicHc]{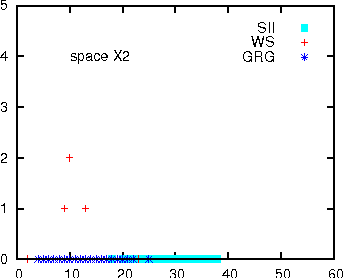} \\
\includegraphics[width=\mosaicMa]{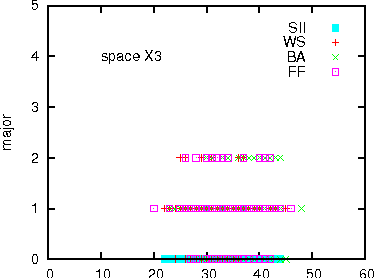} &
\includegraphics[width=\mosaicMb]{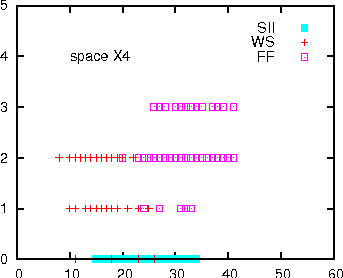} &
\includegraphics[width=\mosaicMc]{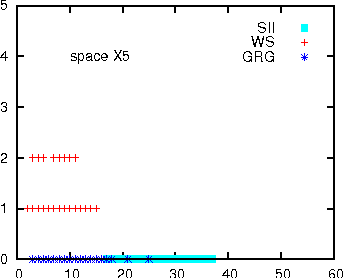} \\
\includegraphics[width=\mosaicBa]{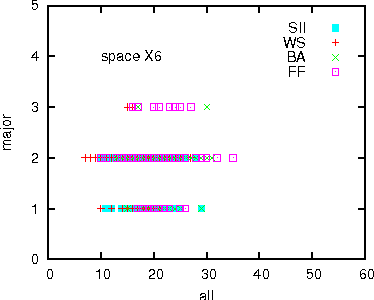} &
\includegraphics[width=\mosaicBb]{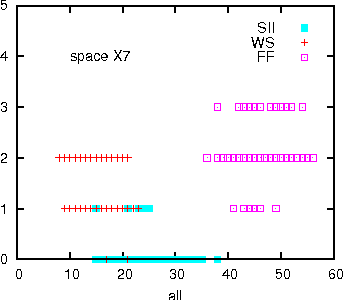} &
\includegraphics[width=\mosaicBc]{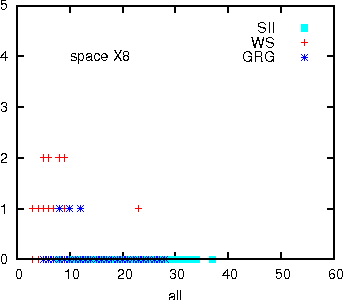} \\
\end{tabular*}
}
\caption{Dynamics $\mathcal{D}^{i}_{opinion}$ of the SIR model over networks $\mathcal{X}^i$ having the same size (lines) and clustering rate (columns)}
\label{fig:OpinionBC_spaceAx}
\end{figure} 

%\paragraph{}
Analysis of simulation results for all subspaces $\mathcal{X}^i$ (fig~\ref{fig:OpinionBC_spaceAx}) reveal a strong sensitivity to the clustering of networks: simulations over networks having a clustering rate of $0.58$ (right column in the figure) do not lead to as many major opinions than networks having a lower clustering rate (left and center columns). These results reveal that denser and more clustered networks require more time to converge towards major opinions. Even if spaces $\mathcal{X}^6$ and $\mathcal{X}^3$ appear to lead to similar results in the figure, a more careful analysis shows that important differences subsist in these spaces, with SII networks leading to less major opinion than others. In general, survival of minor opinions is often facilitated by the core-periphery structure of FF models, while SII networks structured by communities lead to more minor and less major opinions. No general law governs these results, however, as FF may lead either to numerous ($\mathcal{X}^7$, $\mathcal{X}^4$) or few ($\mathcal{X}^0$, $\mathcal{X}^1$) minor opinions depending to the subspace in study. \textit{The possibility to obtain more coherent results by a quantitative definition of network properties is thus not proved yet.}

\subsection{Axelrod\label{indoc:XP_Axelrod}}

\subsubsection{Model}

% The Axelrod Culture Model (ACM) is a simple model that 
% 
% In the ACM, the structure of interactions is classically a regular lattice, with agent being placed over a square array. The agents are entities which may be thought as villages. The model intends to describe the dynamics of culture - as well as TODO. 
% 
% This model is based on the following principles: features are independant, similar agents have an higher probability to interact, 
% 
% As in the original paper, (a) an agent is selected randomly from the population 
% 
% \subsubsection{Indicators}
% 
% %\paragraph{}
% Clusters \citep{samuel_thiriot:bib_sma_simulation:xiao_2009_1}

%\paragraph{}
The Axelrod's model of cultural diffusion \citep{samuel_thiriot:bib_sma_simulation:axelrod_1997_3,samuel_thiriot:bib_sma_simulation:axelrod_1997_4} stands as a famous illustration of agent-based models. This networked model was first proposed on regular lattices, but was recently extended to complex networks \citep{samuel_thiriot:bib_sma_simulation:klemm_2003_1}. In this model, the culture of each agent is formalized as a vector of \textit{cultural features}, each of these features being randomly chosen at initialization from a finite-size set of \textit{cultural traits}. At each step, one agent $i$ is randomly selected in the population (this agent is said to be ``activated'') and one its neighbors $j$ selected at random. The probability for these two agents to actually interact is determined by the similarity of their culture: the more identical values they have for each feature, the more probable the influence is (see~\citep{samuel_thiriot:bib_sma_simulation:axelrod_1997_4} for details). In case of influence, the activated agent changes one of his cultural features (different from the other agent) for the trait of the other agent, thus becoming more similar to this agent. We will arbitrarily use 5 features and 3 traits for our experiments, leading to $3^5 = 243$ possible cultures. 

%\paragraph{}
As the convergence of the model may be very slow, we stop the simulation after a fixed number of steps. In order to enable the comparison of different sizes of networks, this number of steps is related to the number of nodes: $250.N$. As a consequence, whatever the network size, each node will be activated 250 times. Factor 250 was chosen because it is sufficient for the individual cultures to converge to a small set of major cultures. At the end of the simulation, we compute the histogram of cultures against the number of individual who share them. Some simulations may lead to a large number of small cultures without any large culture, while other may lead to a convergence to less than 10 cultures without any minor culture. As both of these indicators may be of use for sociological interpretation of simulation results, we will analyze the dynamics of the Axelrod model on the two-dimensional space having the total number of cultures (whatever the number of agents that share them) in abscissa and the number of major cultures (shared by at least 5\% of the population) in the ordinate axis.  
% Conclusivity of simulation results will be understood as a discrimination in this space, lthat either a maximum or minimal number of major/minor cultures are expected in the model. 

\subsubsection{Simulations over small-worlds}

\begin{figure}[h]
\ifthenelsehtml{
\includegraphics{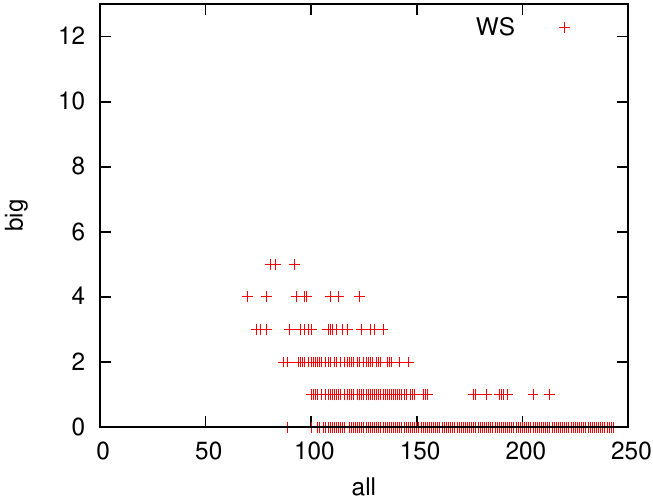}
\includegraphics{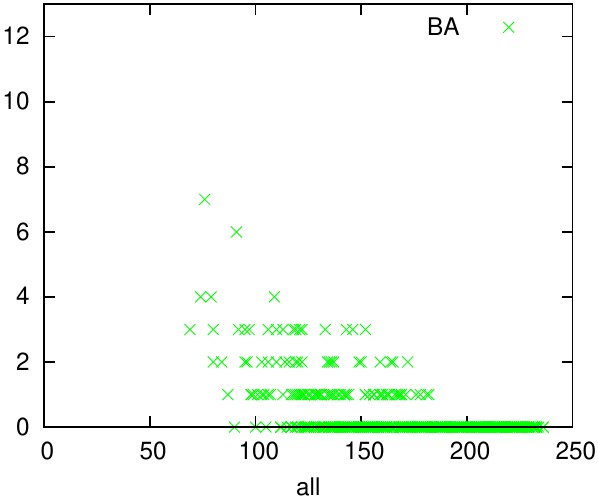}
\includegraphics{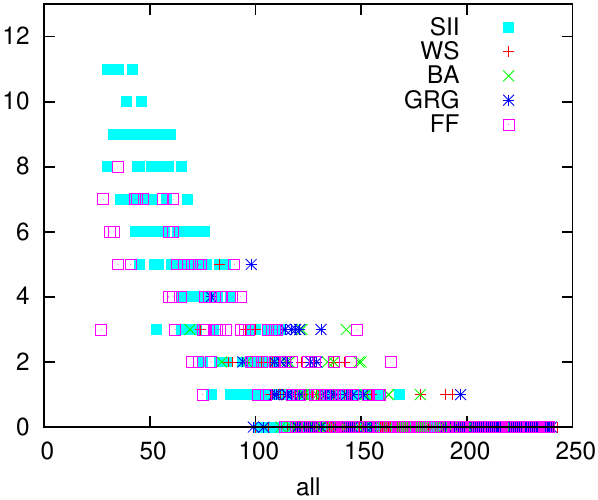}
} {
\includegraphics[height=\hautPrTrois]{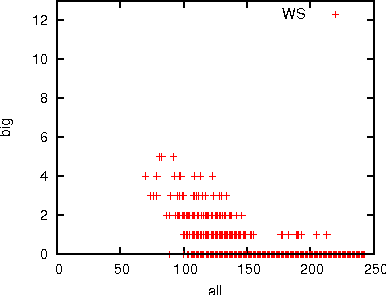}
\includegraphics[height=\hautPrTrois]{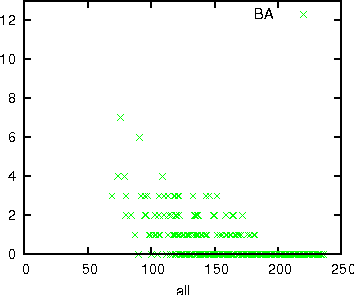}
\includegraphics[height=\hautPrTrois]{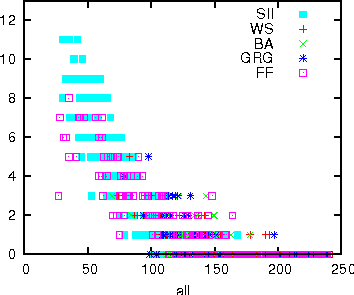}
}
\caption{Simulation results for the Axelrod model ran over \textit{(left)} Watts-Strogatz networks, \textit{(center)} Barab\'asi-Albert networks and \textit{(right)} various small-world networks}
\label{fig:axelrod_all_results}
\end{figure} 

%\paragraph{}
Simulations over Watts-Strogatz and Barab\'asi-Albert networks (figure~\ref{fig:axelrod_all_results} (left\&center)) are quiet similar, the only difference being the slightly higher number of total and big communities in the second case. However, these results are very different from simulation results over the whole space of small-worlds (figure~\ref{fig:axelrod_all_results} (right)), which also contains numerous big communities with few small communities, the total number of big communities being three times the number observed over WS and BA networks. In other words, these experiments prove that \textit{simulations of the Axelrod model of culture using Barab\'asi-Albert and Watts-Strogatz are not representative of the dynamics of the model over small-world networks in general}. Moreover, as for previous models, simulations over small-worlds in general are very scattered. These results do not enable to assess the model validity (whatever the number of major and minor opinions observed in the real population, it corresponds to a possible result of the system) nor to restrict the space of expected dynamics in the population; we then consider these results to be inconclusive.

\subsubsection{Simulations for similar small-world networks $\mathcal{X}^i$}

\begin{figure}[htp]
\ifthenelsehtml{
\begin{tabular*}{\textwidth}{lcr}
\includegraphics{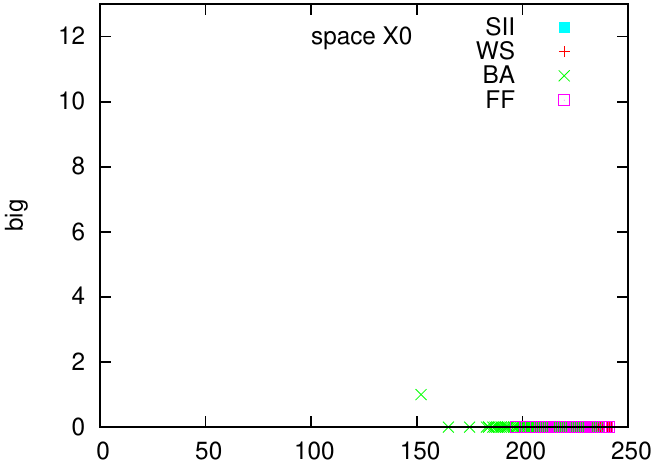} &
\includegraphics{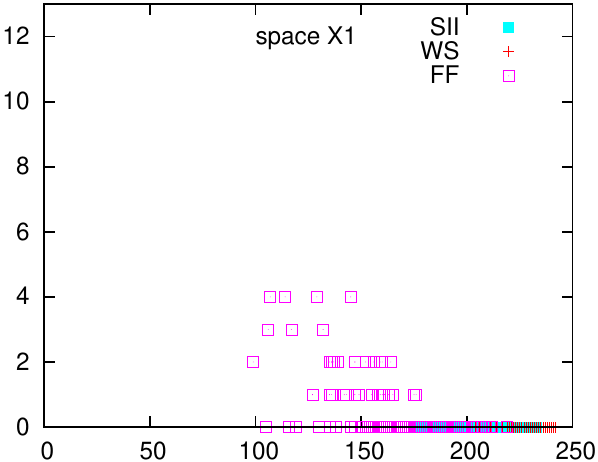} &
\includegraphics{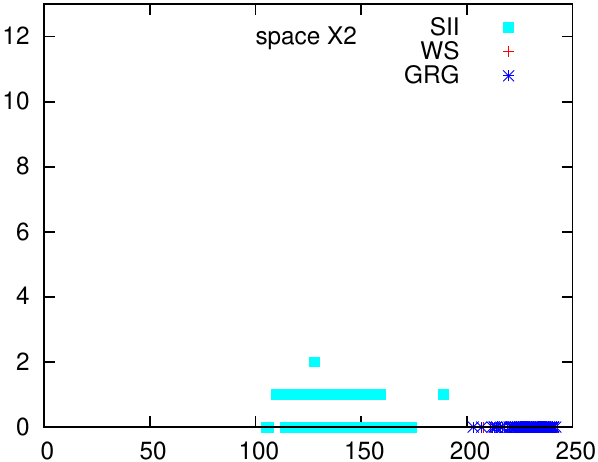} \\
\includegraphics{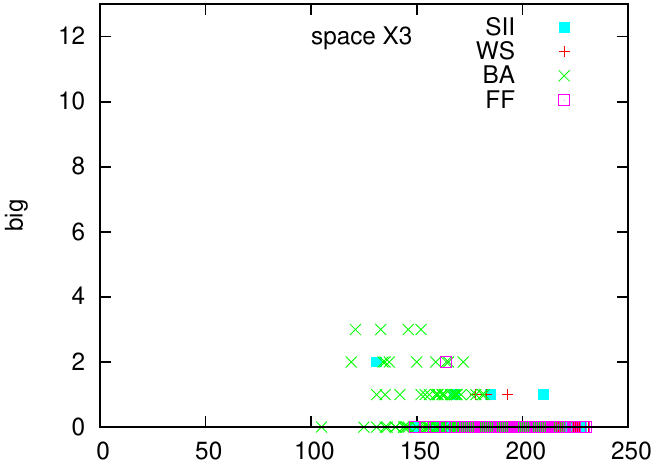} &
\includegraphics{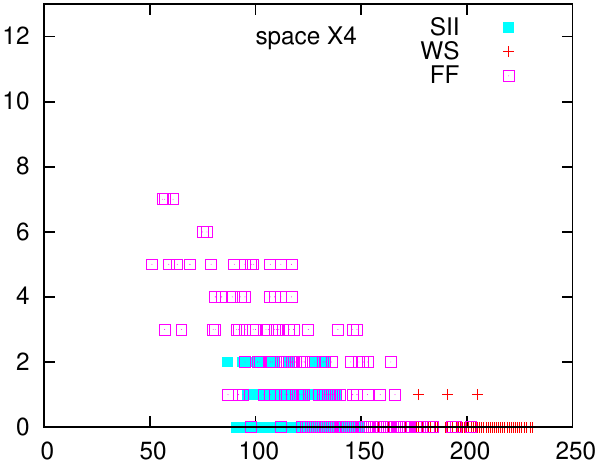} &
\includegraphics{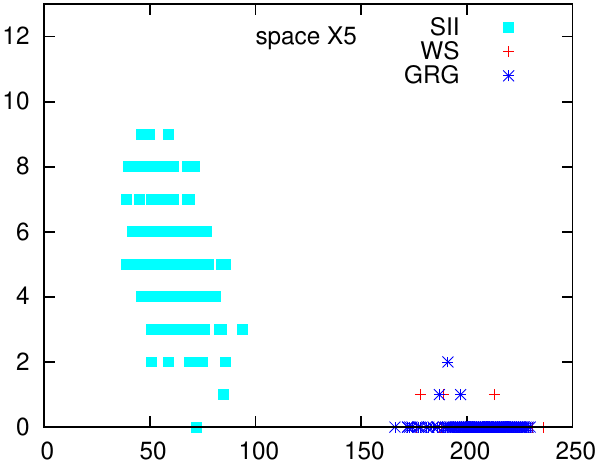} \\
\includegraphics{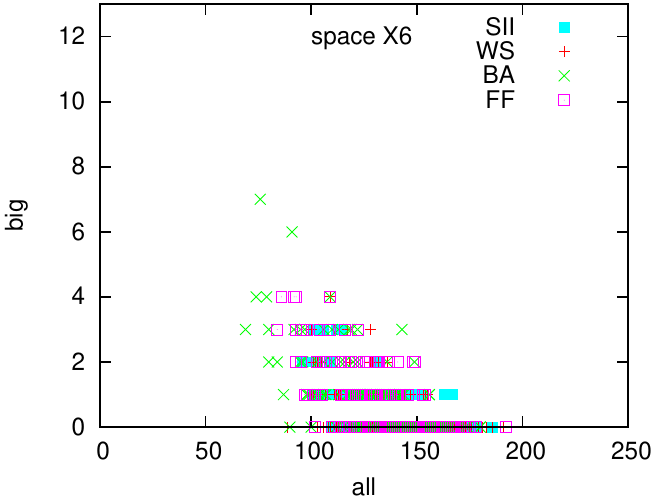} &
\includegraphics{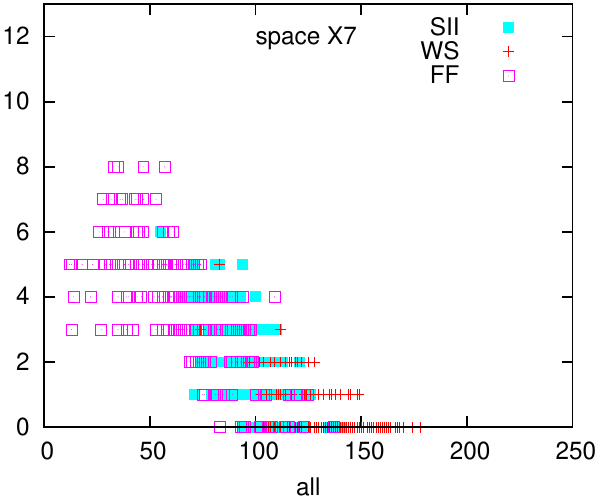} &
\includegraphics{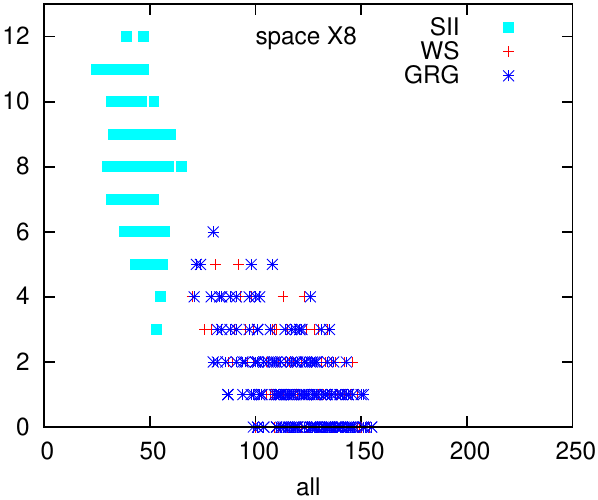} \\
\end{tabular*}
} {
\begin{tabular*}{\textwidth}{lcr}
\includegraphics[width=\mosaicHa]{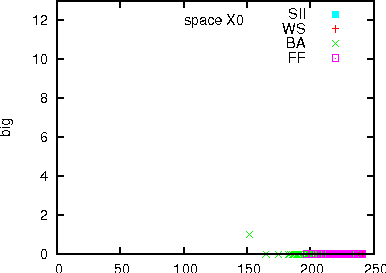} &
\includegraphics[width=\mosaicHb]{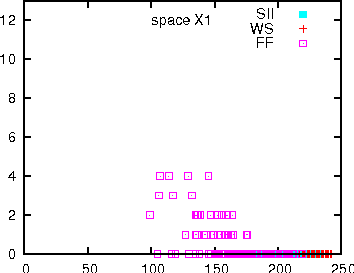} &
\includegraphics[width=\mosaicHc]{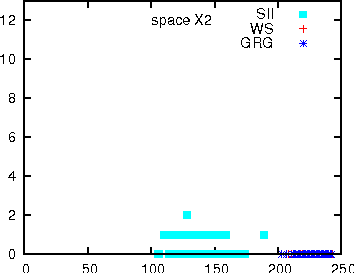} \\
\includegraphics[width=\mosaicMa]{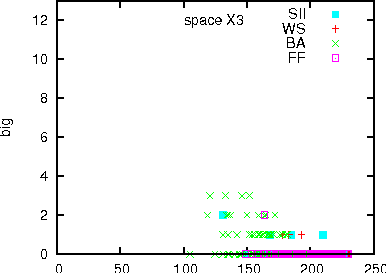} &
\includegraphics[width=\mosaicMb]{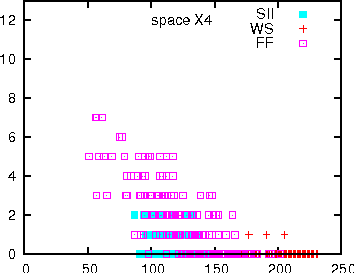} &
\includegraphics[width=\mosaicMc]{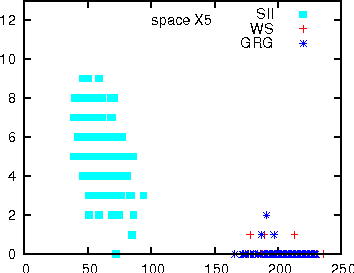} \\
\includegraphics[width=\mosaicBa]{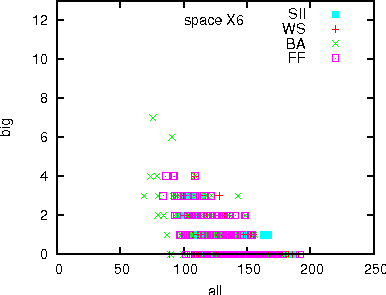} &
\includegraphics[width=\mosaicBb]{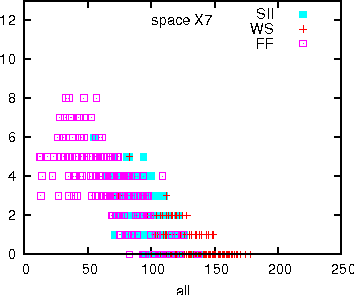} &
\includegraphics[width=\mosaicBc]{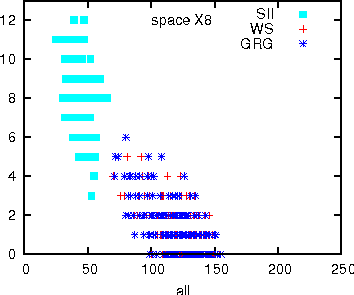} \\
\end{tabular*}
}
\caption{Dynamics $\mathcal{D}_{\text{Axelrod}}^i$ of the Axelrod model for each space $\mathcal{X}^i$. Note that results for WS networks are superposed with those of GRG for spaces $\mathcal{X}^2$, $\mathcal{X}^5$ and $\mathcal{X}^8$.}
\label{fig:Axelrod_spaceAx}
\end{figure} 

%\paragraph{}
Figure~\ref{fig:Axelrod_spaceAx} depicts simulation results for each subspace of networks $\mathcal{X}^i$. Only the space $\mathcal{X}^6$ exhibits similar results whatever the network generator; all the others lead to more or less different results. In spaces $\mathcal{X}^2$, $\mathcal{X}^5$ and $\mathcal{X}^8$ having the higher clustering, the dynamics over SSI suggests that communities lead to numerous big cultures and few minor cultures, while GRG and WS lead to very similar results. In spaces $\mathcal{X}^1$, $\mathcal{X}^4$ and $\mathcal{X}^7$ characterized by an intermediate value of clustering, FF models always lead to more major cultures and less minor cultures than in SII and WS networks. In the left column that contains results for networks having a low clustering, BA appears to facilitate the apparition of major cultures and reduce the number of minor cultures. Given the diversity of results even in these spaces of networks having the same clustering, density, size and average path length, \textit{the refiniability potential of small-world criteria appear to be in-existant.}

% 
% %\paragraph{}
% The comparison of spaces having the same size (horizontally) or clustering rate (vertically) do not suggest any similarity due to one of these properties. Despite of a deeper analysis of these results (against network size, density, clustering, short average path length or combinaisons of them), we did not found any law governing dependencies between these spaces. In other words, given these experiments, nothing prooves that defining more precise values for transivity, density or average path length would lead to more conclusive results. The refinability potential of the criteria invoved in the small-world phenomenon appears quiet low given our experiments.
% 

\section{Discussion\label{indoc:discussion}}

% One problem with the space of possible graphs is that it is, till now, impossible to quantify the size of the networks compliant with several properties in the general case. As a consequence, our experiments only provide some samples from this space of graphs. It is sufficient for us to prove that some models are very sensitive in some spaces, while 

\subsection{Summary}

%\paragraph{}
In this paper, we underlined a potential flaw in common practices in computational simulation: many modelers argue that their simulations make sense because they use plausible networks, but actually study the dynamics of their models over specific samples of these plausible networks. We first identified three possible problematics related to this fact: (i) the possible \textit{lack of representativity of a generator to the class of networks} of interest, (ii) the possible \textit{lack of conclusiveness of simulations over a class of networks} and (iii) the \textit{possibility to refine the criteria of network choice} for reaching (more) conclusive results. We formalized these problematics using the concept of criteria of network choice, spaces of networks and spaces of dynamics. We argued that the criteria for network choice should not  only be based on the plausibility of networks but should also limit the space of plausible networks enough for simulation results to be conclusive. 

%\paragraph{}
We proposed an experimental protocol to tackle these three problematics, and applied it to the space of small-world networks, and the two Watts-Strogatz and Barab\'asi-Albert networks that are often used in agent-based modelling. 
% 
% %\paragraph{}
% The relevance of the indicators defined to track simulation dynamics may be contested; for instance, we used quanitative indicators on one particular setting of each model instead of exploring the behaviour of the model in its space of parameters. Thus the behaviour of the model may appear to be similar wathever the network, the only difference being the value of an epidemic threshold of the time required to convergence. Such a criticism is relevant when building analogical models (TODO ref) that are focused on the study of complex dynamics and are neither intended to be descriptive nor predictive. However, epidemic model \textit{are used} to emit recommandation about vaccination strategies. Even if the simple models of cultural evoluation and opinion dynamics are acknoledged to be too simple to be compared with real processes in a straighforward way \citep{samuel_thiriot:bib_sma_simulation:deffuant_2003_1}, the impact of the structure on them is representative of the sensitivity of more descriptive models. 
% 
% Another relevant critism on this study is shared by any other research: the protocol was built for demonstrating that discripencies exist over networks, while other parameter settings would have led to similar results. As an answer to this remark, this protocol was precisely intended to unveal the potential examples that proove that a network generator cannot be said to be representative of its class of networks, that dynamics over this class of networks is far from TODO.

\subsection{Results for small-world networks}

% %\paragraph{}
% As said in introduction, the \textit{problematic of defining the criteria for the choice of networks was not explicitely formalized}. We proposed a conceptual view of this problematic (\ref{indoc:formal}) by reusing the concepts of spaces of networks, by defining the criteria of network as a set of propositions that define subspace of networks, and by viewing simulations as a projection from the space of networks to a space of dynamics. Using these definitions, we argued that \textit{the criteria for the network choice is not only based on the plausibility of the network}, which actually is not an end in itself, \textit{but also on the coherency of simulation results in the space of dynamics}. 

\subsubsection{Generator representativity: Watts-Strogatz is not small-worlds}

%\paragraph{}
When comparing the dynamics of WS, BA and other network dynamics, it appears that none of these algorithms is representative of the behavior of the model in the whole space of small-world models. Watts-Strogatz networks are only specific examples of small-worlds that lead to specific simulation results. As a consequence, \textit{simulation results (as well as analytical conclusions) obtained with Watts-Strogatz networks should not be extrapolated to the possible dynamics of a real system without further analysis.}  Given these observations, it is of prime importance to keep this representitivity problem in mind when analyzing the results of simulations. The semantic shift from ``Watts-Strogatz networks'' to ``small-world networks'', frequently observed in papers, contributes to the risky assimilation of the specific networks generated by the WS algorithm to the class of small-world networks.

%\paragraph{}
Moreover, the similarity of results observed in some cases (spaces $\mathcal{X}^6$ and $\mathcal{X}^0$ for epidemics dynamics, $\mathcal{X}^3$ for opinion dynamics and $\mathcal{X}^6$ for Axelrod's model) when using Watts-Strogatz and Barab\'asi-Albert networks could confuse modelers, letting them belief that these results are representative of the model behavior, while our experiments clearly reveal important differences when exploring other networks. This similarity between WS and BA results is somewhat ironic, as modelers precisely use them to explore two hypothesis on the possible distribution of degrees in the real networks.

% 
% % More generally, simulation results obtained using one or two specific examples should not be extrapolated to emit recommandations nor for validation purpose without more enquiries over other networks. 
% %\paragraph{}
% Experiments reveal a risk to observe similar results for specific combinaisons of parameters; for instance, space $\mathcal{X}^6$ leads to coherent simulation results wathever the network for opinion dynamics, spaces $\mathcal{X}^6$, $\mathcal{X}^3$ and $\mathcal{X}^0$ for epidemics, $\mathcal{X}^6$ for Culture.
% 
% Refine criteria : $\mathcal{X}^6$ leads to coherent results wathever the network for the three models. 
% 

\subsubsection{Simulation conclusiveness: Small-world is not enough}

%\paragraph{}
As soon as the lack of representativity of a generator to the class of plausible networks is proved, results of simulations should not be studied in the specific case of the network generated by one or two algorithms, but should be interpreted in the whole space of networks that are assumed to be plausible. However, we pointed out that the dynamics over this whole space, which are rarely (if done) explored, may be so scattered that the simulations results would actually be inconclusive. The definition of which result is conclusive (or not) is a complex topic that depends on the model and on the purpose of the modeler. Nevertheless, \textit{the results obtained here are arguably inconclusive}: epidemic dynamics (cf.~\ref{indoc:sir_whole}) predict any extend of the epidemic, thus being as insightful that a purely random hypothesis, while opinion dynamics and culture dynamics lead to results so disparate that the model could fit any observation from the field without restricting much the space of predicted states. These results strongly suggest that \textit{the criteria defined by the small-world phenomenon don't constrainst the dynamics enough for simulation results to be conclusive}. 

%\paragraph{}
Claiming the plausibility of results because they are based on small-world networks is as relevant as claiming their plausibility because of the use of a network rather than assortative mixing: \textit{the ``small-world'' criteria for network choice is necessary - as it was proved to change the behavior of models - but not sufficient to obtain conclusive results.} As discussed later, this problem extends to any exploration of a space of networks: criteria defined to choose the space of networks should not only be based on plausibility but also on their ability to lead to conclusive results.

\subsubsection{Criteria refinability: Small-world cannot be enhanced}

%\paragraph{}
The very question of refinability of criteria is: can we improve our simulations results by measuring the values of the properties of real networks~? Intuitively, refining these criteria is expected to improve the coherency of simulation results: the more precise the criteria $E^{\text{plausible}}$ are, the smaller the corresponding space of networks $\mathcal{X}^{\text{plausible}}$ , the more stable the simulation results $D^{\text{plausible}}$ . For instance, modelers interested in epidemics often parameter network generators for them to comply with a plausible average degree of connectivity (e.g. \citep{samuel_thiriot:bib_sma_simulation:small_2005_1}). However, simulation of three models prove that \textit{even networks having similar clustering, density, size and average path length may lead to qualitatively and quantitatively different results}. Even if, in specific cases, specific samples of the space of small-world networks are shown to lead to similar results, simulation conclusiveness is not guaranteed by refining criteria of network choice. This result confirms the usefulness of taking other characteristics of networks into account, like the distribution of degree, assortativity, diameter and other indicators. However, the conclusiveness of simulation results obtained using these novel criteria should also be challenged; for instance, even if FF and BA have both a fat-tailed distribution of degree, our experiments show that they lead to different results.

\subsection{Implications for computational simulation}

\subsubsection{Explore the space of plausible networks to avoid the representativity bias}

%\paragraph{}
Beyond the specific cases of the Watts-Strogatz and Barab\'asi-Albert generators, our results suggest that simulations over generated artificial networks are not sufficient to study the possible behavior of models over what we trust to be plausible networks. This may constitute a fundamental limitation for computational simulation in its whole. \textit{Whatever the network generator you choose} (even if it is a ad-hoc algorithm), \textit{results may not be representative to the possible dynamics} (over the networks you belief to be plausible). Worst, as no general method permits to determine whether the entire space of networks $\mathcal{X}^{\text{plausible}}$ was extensively explored or not, the relevance of simulation results over artificial networks will always remain questionable. Nevertheless, the bias induced by the choice of specific network generators may be reduced by a more extensive exploration of the space of networks that are assumed to be plausible.
% Note that similar problems may be identified using a real network as the structure of interactions (as done in \citep{samuel_thiriot:bib_sma_simulation:kempe_2003_1} or \citep{samuel_thiriot:bib_sma_simulation:cointet_2007_2}), one unique network probably being not be representative of the dynamics occuring over the various possible structure of interactions. 

%\paragraph{}
The \textit{use of several network generators} having different properties and combinations of properties, as done in this paper, could help to explore the behavior of a model over the space of plausible networks. \textit{Rewiring algorithms} \citep{samuel_thiriot:bib_sma_simulation:kawachi_2004_1,samuel_thiriot:bib_sma_simulation:watts_1998_1,samuel_thiriot:bib_sma_simulation:molloy_1995_1} \textit{may also constitute valuable tools to explore this space}, as they often guarantee the preservation of several characteristics of the original networks (e.g. size or distribution of degree) while introducing some noise in the structure (see the previous applications \citep{samuel_thiriot:bib_sma_simulation:cointet_2007_1,samuel_thiriot:bib_sma_simulation:dekker_2007_1}). The use of other generators and of rewiring is limited by the cost of such an experiment for the modeler, who would have to implement other network generators (development cost) and drive more simulations (computational cost). Moreover, this process should not make more difficult the communication of results, even if the use of several generators requires the description of these algorithms to enable replication.
 
%\paragraph{}
In order to limit the development cost, the \textit{development of software or libraries interoperable with simulation plateforms} may be of help. Several network generators that explore different subspaces of small-worlds could be identified (for instance, testing both WS and GRG networks appear to be useless in the present experiments, as both lead to very close results), thus limiting the number of algorithms to implement, facilitating communication of simulation results and saving computational time. The simulation cost can be reduced by \textit{avoiding the systematic generation of networks}, which (along with the analysis of their statistical properties) is very costly. To do so, further research should explore the possibility to \textit{create samples of the network generated by one algorithm that are representative of the generated networks} (for instance, generating thousands of Watts-Strogatz networks with a given parameter setting is probably enough to study the dynamics of a model among the various configurations created by this generator). If such a sampling is possible, a library of networks (large number of networks compliant with given properties) could be built and shared with the whole research community, thus enabling replication and preserving communicability. A good sampling of a space of networks of interests, would they be small-world or scale-free, could thus be tested at low cost, and be easily communicated by only citing the reference of the set of networks involved in simulations.

\subsubsection{Identify other criteria to enhance conclusiveness}

%\paragraph{}
As observed here for small-world networks, the exploration of the space of networks we assume to be plausible may reveal inconclusive results. However, we are not interested in the specific case of the artificial networks generated by one algorithm, but rather on the dynamics of our models over the various possible networks that may exist in the real society (i.e. the plausible networks).

Inconclusiveness may be due to a systemic cause, that is the social phenomenon studied is unpredictable in itself; however, as the identification of characteristics of real networks is still recent, it is far much probable that other characteristics of real networks could be included as criteria of network choice, thus improving the benefit of simulations. Several novel properties were already identified for real networks \citep{samuel_thiriot:bib_sma_simulation:jackson_2008_1,samuel_thiriot:bib_sma_simulation:newman_2006_1}, including a skewed distribution of degree, a core-periphery structure, the existence of communities \citep{samuel_thiriot:bib_sma_simulation:girvan_2002_1,samuel_thiriot:bib_sma_simulation:palla_2005_1}, positive assortativity of degree, spectra of networks~\citep{samuel_thiriot:bib_sma_simulation:farkas_2001_1}, clumpiness \citep{samuel_thiriot:bib_sma_simulation:snijders_2006_1} or node centrality \citep{samuel_thiriot:bib_psycho:borgatti_2005_1}. However, as demonstrated in this paper, these indicators should not be chosen only for their ability to discriminate the plausible networks to the implausible ones, but also because they reduce the space of supported dynamics enough for simulation to be conclusive; such a checking may be based on the protocol detailed in this paper.
% Note that the stream of complex networks focused on the identification of the properties shared by both social and artificial networks, while we are interested in the actual characteristics of the networks of interactions.
% 
% %\paragraph{}
% Network generators were already proposed to mimic the characteristics recently observed in real networks (Scale free networks with high clustering \citep{samuel_thiriot:bib_sma_simulation:holme_2002_1,samuel_thiriot:bib_sma_simulation:klemm_2002_1}, or uncorrelated \citep{samuel_thiriot:bib_sma_simulation:catanzaro_2005_1}). 

% This study may even be useful \textit{before} proving the generality of the new characteristics, in order to focus on the measure of this characteristic only 

\subsubsection{Explore novel approaches}

%\paragraph{}
This paper may be viewed as another criticism of the wide use of artificial networks for social simulation. Till now, the main criticism is their \textit{lack of plausibility},  \citep{samuel_thiriot:bib_sma_simulation:roth_2007_1,samuel_thiriot:bib_sma_simulation:cointet_2007_1,samuel_thiriot:bib_sma_simulation:pujol_2005_1}. From a methodological viewpoint, the network generators developed in the stream of complex networks are simple algorithms that aim only to reproduce some precise characteristics observed in real networks, that are \textit{used out of their initial purpose and may be irrelevant for the peculiar needs of social simulation}. 

% Both these arguments make intuitively sense, but 

%\paragraph{}

Several alternative approaches are already explored for the generation of networks:
\begin{itemize}
\item \textit{Create the networks using the available part of the real structure of interactions}. This approach is notably used to simulate epidemics using transportation data \citep{samuel_thiriot:bib_sma_simulation:ferguson_2006_1,samuel_thiriot:bib_sma_simulation:colizza_2007_1,samuel_thiriot:bib_sma_simulation:germann_2006_1}. This approach, however, is only usable when data is already available for the interactions involved in the model, which is rarely the case in practice. 

\item Several researchers explore the \textit{generation of networks from local and plausible behaviors}, using concepts like spatial or social distance \citep{samuel_thiriot:bib_sma_simulation:wong_2006_1} or social circles \citep{samuel_thiriot:bib_sma_simulation:hamill_2009_1}. Nevertheless, even if these models rely on plausible rules of network generation, these ones cannot be easily compared with real networks (unless using statistical properties that were shown here to be too permissive), thus allowing uncertainty to persist about their relevance. 
% Moreover, these ad-hoc algorithms could suffer the same problems than those detailed here for artificial networks: these constrainsts on networks being permissive enough for the corresponding space of networks to remain inconclusive. 

\item The very last approach consists in \textit{using data available on the general conditions of interaction to generate the network}. For instance, the network may be generated from census data (\citep{samuel_thiriot:bib_sma_simulation:eubank_2004_1}), individual interviews \citep{samuel_thiriot:bib_sma_simulation:amblard_2001_1} or household data \citep{samuel_thiriot:bib_sma_simulation:meyers_2005_1}. However, these methods lack genericity (thus complicating the communicability of simulation results) and require the development of ad-hoc generative algorithms, therefore failing to constitute a tool easily usable by the numerous newcomers in agent-based modelling. 
% The representativity of each specific algorithm based on such data should also be checked for them to constitue valuable tools.

\end{itemize}

%\paragraph{}
The ideal network generator should integrate the benefits of all of these approaches, by \textit{using available observations from the field as parameters}, and using \textit{plausible local rules} to generate the network of interactions, while remaining tools usable without deep skills in computer science. Recent approaches developed specifically for social simulation (Nota Bene: removed for blind review 
% \citep{samuel_thiriot:bib_perso:thiriot_2008_3}
) just start to investigate this promising lead. 
% The parameters' setting of such a network generator would be more numerous and time-consuming than simple algorithm; this 

\subsection{Conclusion}

%\paragraph{}
% Note that this entire paper assumes the use of simple networks (unweighted, uniplex, undirected). The use of multiplex networks or weighted networks lead to even more 

Social networks constitute a complex problematic which studies an object that \textit{cannot be observed from the field} while having a \textit{dramatic influence on simulation results}. Moreover, this object is \textit{used as a parameter for most agent-based simulations}. We first highlighted the lack of representativity of artificial networks generated by one algorithm to the space of plausible networks, and proposed to rely on libraries of networks as testbeds. However, our experiments also suggest that constrainsts based on the statistical properties of networks may be to permissive to obtain conclusive results, even when refining criteria for network choice by using quantitative values. \textit{As inconclusive results reduce dramatically the benefits of social simulation for decision-making, the proposal of new approaches for the description of interaction networks constitute a topic of first importance for social simulation in its whole.}

\section*{Acknowledgements}

%\paragraph{}
% (Nota Bene: removed for blind review)
Thanks to Edmund Chattoe-Brown, Camille Roth and Christophe Sibertin-Blanc for insightful comments on a previous release of this paper. I would like to thank Frederic Amblard for insightful recommendations on the experimental settings used for this research. I also would like to thank Matthieu Latapy and Edmund Chattoe-Brown for previous discussions about networks. This work was partly funded by the French foundation for science RTRA STAE (R\'eseau Technologie de Recherche Avanc\'ee pour les Sciences et Technologies de l'A\'eronautique et de l'Espace) in the frame of the MAELIA project.

\bibliography{../../commons/biblio_bibtex/bib_customer_value.bib,../../commons/biblio_bibtex/bib_customer_value_private.bib,../../commons/biblio_bibtex/bib_sma,../../commons/biblio_bibtex/bib_sma_organisation,../../commons/biblio_bibtex/bib_decision_psychology,../../commons/biblio_bibtex/bib_modelsimu,../../commons/biblio_bibtex/bib_model_decision,../../commons/biblio_bibtex/bib_psycho,../../commons/biblio_bibtex/bib_sma_simulation,../../commons/biblio_bibtex/bib_ethologie,../../commons/biblio_bibtex/bib_ia_forte,../../commons/biblio_bibtex/bib_general,../../commons/biblio_bibtex/bib_communications_privees,../../commons/biblio_bibtex/bib_perso,../../commons/biblio_bibtex/bib_apply_decision,../../commons/biblio_bibtex/bib_software}
% 
% 
% \section*{Appendix A}
% 
% 
% \begin{figure}[htp]
% \ifthenelsehtml{
% \includegraphics{imagesDif/allNetworks2D_density_transitivity_spaceA}
% \includegraphics{imagesDif/allNetworks2D_length_transitivity_spaceA}
% \includegraphics{imagesDif/allNetworks2D_size_transitivity_spaceA}
% } {
% \includegraphics[width=\textwidth]{imagesDif/allNetworks2D_density_transitivity_spaceA}
% \includegraphics[width=\textwidth]{imagesDif/allNetworks2D_length_transitivity_spaceA}
% \includegraphics[width=\textwidth]{imagesDif/allNetworks2D_size_transitivity_spaceA}
% }
% \caption{Networks generated in space A projected given their transitivity and \textit{(top)} density, \textit{(middle)} average path length and \textit{(bottom)} size.}
% \label{fig:all_networks_2_trans_A}
% \end{figure} 
% 
% 

\end{document}